\documentclass[a4paper,twocolumn,10pt,accepted=2019-07-18]{quantumarticle}
\pdfoutput=1
\usepackage[utf8]{inputenc}
\usepackage[english]{babel}
\usepackage[T1]{fontenc}
\usepackage{amsmath}
\usepackage{hyperref}
\usepackage{bm,bbm,bbold,amssymb}
\usepackage[numbers,sort&compress]{natbib}
\usepackage{amsthm} 
\usepackage{enumerate} 

\newcommand{\tr}{\mathrm{tr}}
\def\bra#1{\langle{#1}|}
\def\ket#1{|{#1}\rangle}

\newcommand{\ketbra}[2]{\ket{#1}\!\bra{#2}}

\def\BraVert{\egroup\,\mid\,\bgroup}

\newcommand{\sgn}{\textrm{sgn}}

\begin{document}

\title{Tight, robust, and feasible quantum speed limits for open dynamics}

\author{Francesco Campaioli}
\email{francesco.campaioli@monash.edu}
\affiliation{School of Physics and Astronomy, Monash University, Clayton, Victoria 3800, Australia}

\author{Felix A. Pollock}
\affiliation{School of Physics and Astronomy, Monash University, Clayton, Victoria 3800, Australia}

\author{Kavan Modi}
\affiliation{School of Physics and Astronomy, Monash University, Clayton, Victoria 3800, Australia}

\date{26 July 2019}

\begin{abstract}
Starting from a geometric perspective, we derive a quantum speed limit for arbitrary open quantum evolution, which could be Markovian or non-Markovian, providing a fundamental bound on the time taken for the most general quantum dynamics. Our methods rely on measuring angles and distances between (mixed) states represented as generalized Bloch vectors. We study the properties of our bound and present its form for closed and open evolution, with the latter in both Lindblad form and in terms of a memory kernel. Our speed limit is provably robust under composition and mixing, features that largely improve the effectiveness of quantum speed limits for open evolution of mixed states. We also demonstrate that our bound is easier to compute and measure than other quantum speed limits for open evolution, and that it is tighter than the previous bounds for almost all open processes. Finally, we discuss the  usefulness of quantum speed limits and their impact in current research.
\end{abstract} 
\maketitle
\makeatletter
\section{Introduction}
\emph{Quantum speed limits} (QSLs) set a lower bound on the time required for a quantum system to evolve between two given states~\cite{Mandelstam1945,Margolus1998,Deffner2013,Deffner2017}. Such bounds are typically applied to estimate the speed of computational gates~\cite{Lloyd2000,Giovannetti2003a}, the precision in quantum metrology~\cite{Alipour2014,Giovannetti2011,Chin2012,Demkowicz-Dobrzanski2012,Chenu2017}, the performance in quantum optimal control~\cite{Reich2012,Caneva2009,DelCampo2012,Hegerfeldt2013,Murphy2010,An2016,Deffner2017,Campbell2017,Funo2017a}, and the charging power 
in quantum thermodynamics~\cite{Campaioli2017, Campaioli2018a}. Besides their practical relevance, speed limit bounds stand as a fundamental result for both classical and quantum systems~\cite{Okuyama2018, Shanahan2018}, providing an operational interpretation of the largely discussed time-energy uncertainty relations~\cite{Deffner2017}. For these reasons, they have received particular attention from the quantum information community in recent years~\cite{Kupferman2008, Uzdin2012,Santos2015,Santos2016, Goold2016, Uzdin2016,Mondal2016, Mondal2016b, Mirkin2016, Pires2016, Marvian2016, Friis2016, Epstein2017, Ektesabi2017, Russell2017, Garcia-Pintos2018, Berrada2018, Santos2018, Hu2018, Volkoff2018}.

The typical blueprint for constructing QSL bounds involves estimating the minimal evolution time $\tau$ as the ratio between some distance between states and the average \emph{speed} induced by the generator of the evolution~\cite{Pires2016, Deffner2017}. For example, when unitary evolution of pure states is considered, an achievable QSL is given by $\tau\geq d_{FS}/\overline{\Delta E}$, where we have set $\hbar \equiv 1$, $d_{FS}$ is the Fubini-Study distance between the initial and final state~\cite{Fubini1904, Study1905, Bengtsson2008}, and $\overline{\Delta E}$ is the time-averaged standard deviation of the Hamiltonian $H$, which plays the role of the average speed~\cite{Mandelstam1945, Levitin2009}. In more general cases, such as for unitary evolution of mixed states or open (Markovian or non-Markovian) dynamics, such as those in Refs.~\cite{Deffner2013,Deffner2013b, DelCampo2013, Sun2015, Pires2016, Mondal2016, Mondal2016b}, QSLs are generally loose, and an attainable bound is not known~\footnote{Except for the case of qubits~\cite{Campaioli2018}.}. Moreover, the tightest of these bounds are difficult to compute or measure, requiring the diagonalization of initial and final states. 

In this Article we propose a geometrically motivated QSL for open quantum evolutions and demonstrate, analytically and numerically, its superiority over the tightest of known QSLs for almost all open processes. Namely, we show our bound's performance discussing two aspects of QSL: feasibility and tightness.

The feasibility of a bound is quantified in terms of the computational or experimental resources requisite to evaluate or measure the bound.~\cite{Deffner2017, Campaioli2018}. Bounds that require the evaluation of complicated functions of the states or the generators of the evolution are less feasibile, and thus less useful, than otherwise equally performing bounds that are easier to compute or experimentally measure. The distance term in many QSLs requires the square root of the initial and final states, thus the solution of the eigenvalue problem for the initial state $\rho$ and final state $\sigma$~\cite{Deffner2013, Deffner2013b, Pires2016, Mondal2016}. In contrast, the bounds that only involve the evaluation of the overlap $\tr[\rho\sigma]$~\cite{DelCampo2013, Sun2015, Campaioli2018}, including the one that we introduce here, are much easier to compute and measure~\cite{Keyl2001, Ekert2002, Mondal2016b}.
Aside from the distance, the other important feature of QSL bounds is the average speed term, discussed above, that depends on the orbit of the evolution~\cite{Bengtsson2008, Russell2014a, Russell2017, Mondal2016,Mondal2016b, Pires2016, Campaioli2018}. A common criticism of QSLs is that calculating these bounds becomes as hard as solving the dynamical problem, reducing their relevance to a mere curiosity. We overcome this limitation by providing an operational procedure to experimentally evaluate the speed term for any type of process, and go on to discuss the purpose of QSLs in this context.

The tightness of QSLs, which represents how precisely they bound the actual minimal time of evolution, becomes a problem as soon as we move away from the case of unitary evolution of pure states~\cite{Levitin2009}\, which is, in practice, always an idealized description. We will show below that the available bounds for the general case of open evolution of mixed states are rather loose. Their performance gets worse as increasingly mixed states are considered -- which constitutes a major issue for the effectiveness of QSLs.
This looseness is often a consequence of the choice of the distance used to derive the bounds: Different distances result in different speed limits, and a suitable choice that reflects the features of the considered evolution is the key to  performance~\cite{Bengtsson2008, Pires2016, Campaioli2018}. In this Article, we directly address this issue, deriving a bound that is provably robust under mixing, vastly improving the effectiveness of QSLs.

The present Article strongly complements the findings of Ref.~\cite{Pires2016}, where the authors used geometric arguments to obtain an infinite family of distances and their corresponding QSL bounds. While their result firmly and rigorously establishes the mathematical framework for a certain class of QSLs, it leaves open the problem of finding a distance that leads to a QSL that is tight and feasible. We do exactly this by uncovering a distance measure on quantum states, which is based on the geometry of the space of density operators, leading to a QSL that is both tight and feasible for almost all states and processes.

\section{QSL for geometric distance}
Usually, to generalize QSL for when the initial and final states are not pure, the Fubini-Study distance, mentioned in the introduction, is replaced by the Bures distance~\cite{Wootters1981, Deffner2013}:
\begin{gather}
\label{eq:Bures}
   B(\rho,\sigma):=\arccos\big(F(\rho,\sigma)\big),
\end{gather}
where $F(\rho,\sigma) := \tr\big[\sqrt{\sqrt{\rho}\sigma\sqrt{\rho}}\big]$ is the quantum fidelity. In Ref.~\cite{Campaioli2018} we showed that for unitary evolution of mixed quantum states, the corresponding QSL can be extremely loose. This looseness can be attributed to certain feature of the Bures metric, which led us to conclude that a more apt metric for the space of mixed states is desirable.

To remedy this problem, we exploited the generalized Bloch representation~\cite{Byrd2003}, for which every mixed state $\rho$ is associated to a real vector $\bm{r}$, known as a generalized Bloch vector (GBV). By noticing that unitary dynamics of the system preserves the radius of such vectors $\bm{r}$ (which is directly related to the purity of $\rho$), we introduced a geometric distance between states, given by the angle $\Theta$ between their GBVs (discussed in details below). The QSL derived from this distance is provably attainable for the case of qubits, and tighter than the traditional QSL for almost all states in the case of higher dimensions~\cite{Campaioli2018}. In the present Article, driven by this geometric consideration, we generalize the distance $\Theta$ to derive a QSL for arbitrary open quantum processes that outperforms the bounds given in Refs.~\cite{Deffner2013b, DelCampo2013, Sun2015, Deffner2013b, Mondal2016}.

We consider a $d$-dimensional system $S$, where $d=\dim\mathcal{H}_S$, coupled to its environment $E$ (with total Hilbert space $\mathcal{H} = \mathcal{H}_S\otimes\mathcal{H}_E$) and denote its physical state space of positive, unit trace density operators by $\mathcal{S}(\mathcal{H}_S)$.
A state $\rho\in\mathcal{S}(\mathcal{H}_S)$ of the system can be expressed as
\begin{gather}
    \label{eq:generalized_bloch_vetor}
    \rho=\frac{\mathbb{1}+c\;\bm{r}\cdot\bm{\Lambda}}{d},
\end{gather}
where $c=\sqrt{d(d-1)/2}$, given an operator basis $\{\Lambda_a\}$ satisfying $\tr[\Lambda_a\Lambda_b]=2\delta_{ab}$ and $\tr[\Lambda_a]=0$.
The generalized Bloch vector $\bm{r}$ is a vector in a $(d^2-1)$-dimensional real vector space, equipped with the standard Euclidean norm  $\lVert \bm{r} \rVert_2=\sqrt{\sum_i r_i^2}$~\cite{Byrd2003}.

We would like to measure the distance between two states $\rho \leftrightarrow \bm{r}$ and $\sigma \leftrightarrow \bm{s}$ using the length of the shortest path through $\mathcal{S}(\mathcal{H}_S)$ that connects $\rho$ and $\sigma$. However, solving this geodesic problem is, in general, a hard task, since the state space for $d>2$ is a \emph{complicated} subset of the $(d^2-1)$-ball containing all (sub-)unit vectors $\bm{r}$~\cite{Byrd2003,Bengtsson2008}. Our approach will be to simplify this problem by lower bounding this distance by the length of the well-known geodesics of this ball. While this leads to a QSL that intrinsically underestimates the actual optimal evolution time, we will see that it provides a significant improvement over other bounds in the literature.
\begin{figure}
    \centering
    \includegraphics[width=0.47\textwidth]{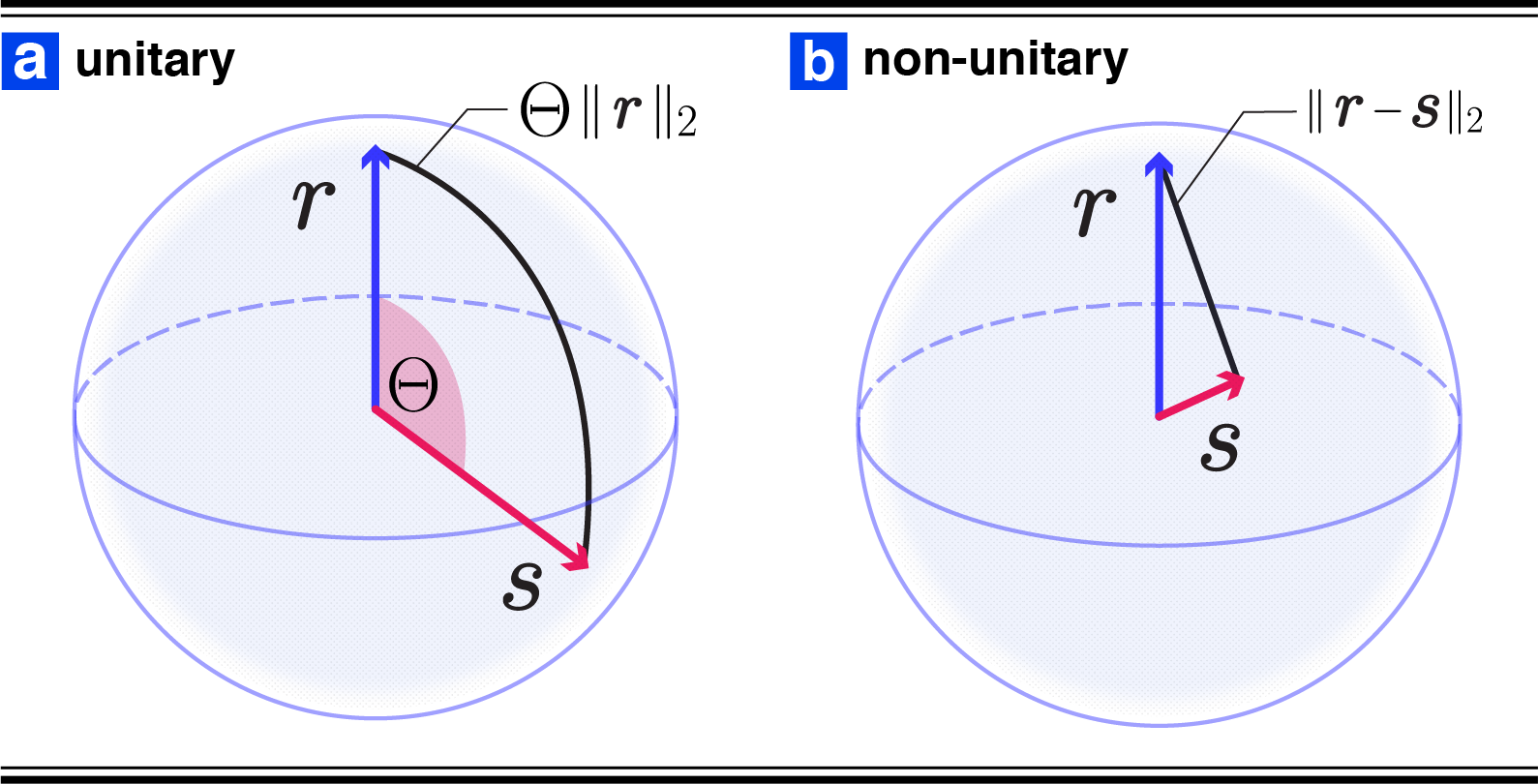}
    \caption{\textbf{Distance on the space of states.} Mixed states $\rho$ and $\sigma$ are represented by their generalized Bloch vectors $\bm{r}$ and $\bm{s}$. In order to simplify the evaluation of the distance $D$ between states we approximate the state space with such $(d^2-1)$-dimensional ball. Accordingly, for the case of unitary evolution ({\fontfamily{phv}\selectfont \textbf{a}}) we choose to measure the distance between $\rho$ and $\sigma$ as the length of the arc of great circle that connects $\bm{r}$ and $\bm{s}$, given by the product between the generalized Bloch angle $\Theta$ between the two vectors, and their length $\lVert \bm{r}\rVert_2$. For the case of arbitrary non-unitary evolution ({\fontfamily{phv}\selectfont \textbf{b}}), we use, instead, the norm of the displacement vector $\bm{r}-\bm{s}$, since the shortest path that connects the two states is given by the straight line (hyper-spheres here schematically represented as 2-spheres).}
    \label{fig:distances}
\end{figure}

For general evolutions, including \emph{non-unitary} open evolution, the length of the GBV is allowed to change. Here, the natural choice for the geodesic is with respect to the Euclidean distance, which is just the straight line between $\bm{r}$ and $\bm{s}$ (see Fig.~\ref{fig:distances} {\fontfamily{phv}\selectfont \textbf{b}}), whose length is given by 
\begin{gather}
    \label{eq:distance}
    D(\rho,\sigma)=\lVert \bm{r}-\bm{s}\rVert_2.
\end{gather}
From this distance we derive our speed limit, following the same geometric argument used in Refs.~\cite{Pires2016,Campaioli2018}, and other QSL derivations. 

By definition, the distance $D(\rho,\sigma)$ is smaller than or equal to the length $L[\gamma_{\rho}^{\sigma}]=\int_0^\tau D(\rho_{t+dt},\rho_t)$ of any path $\gamma_{\rho}^{\sigma}= [\rho_t]_{t=0}^\tau$, generated by some dynamical process, that connects $\rho=\rho_0$ and $\sigma=\rho_\tau$. We evaluate the infinitesimal distance $D(\rho_{t+dt},\rho_t)$ and rearrange to obtain $\tau \geq D(\rho,\sigma)/\overline{\lVert \dot{\bm{r}}_t\rVert}_2$,
where $\overline{f(t)}=\int_0^\tau dt\: f(t)/\tau$ is the average of $f(t)$ along the orbit parameterized by $t\in[0,\tau]$. Expressing $\lVert \bm{r}-\bm{s}\rVert_2$ and $\lVert \dot{\bm{r}}_t\rVert_2$ in terms of $\rho$ and $\sigma$,
we obtain the bound $T_{D}$,
\begin{align}
    \label{eq:speed_limit_arbitrary}
    & \tau \geq T_{D} = \frac{\lVert \rho - \sigma \rVert}{\overline{ \lVert \dot{\rho}_t\rVert}},
\end{align}
where the Hilbert-Schmidt norm $\lVert X \rVert = \sqrt{\tr[X^\dagger X]}$ of an operator $X$ arises as a consequence of equipping the space of GBVs with the standard Euclidean norm~\cite{Bengtsson2008}.

Despite its surprisingly simple form, reminiscent of kinematic equations, the bound in Eq.~\eqref{eq:speed_limit_arbitrary} originates from a precise geometric approach and encompasses the fundamental features of previous QSL bounds, including the orbit dependent term $\overline{ \lVert \dot{\rho}_t\rVert}$, which will be referred to as \emph{speed}, or \emph{strength of the generator}, that appears, under various guises, in the bounds of Refs.~\cite{Deffner2013, Deffner2013b, Taddei2013, DelCampo2013, Sun2015, Pires2016, Mondal2016, Mondal2016b} (see Appendix~\ref{a:derivation} for details about distance $D$ and the derivation of bound $T_D$).

Below we elaborate on several key properties of the geometric QSL given in the last equation. First, we show that this QSL is robust. Next, we give the exact form of the denominator for several important classes of dynamics. Finally, we discuss the superiority of this QSL by showing its feasibility and tightness.

\section{Robustness of geometric QSL}

\subsection{Robustness under composition}
\label{s:robustness_composition}

Our bound is robust in two important ways. First, it is well-known that the Hilbert-Schmidt norm is generally not contractive for CPTP dynamics~\cite{Perez-Garcia2006}.  This means that the distance between $\rho$ and $\sigma$ changes drastically when an ancillary system $\alpha$ is introduced trivially, i.e., $\rho \to \rho \otimes \alpha$ and $\sigma \to \sigma \otimes \alpha$. Then we have $ \| \rho\otimes \alpha-\sigma \otimes\alpha \|  =  \| \rho-\sigma \| \cdot \|\alpha\|$, where the last term is the purity of the ancillary system. Simply by introducing an ancilla that is not pure decreases the original distance by the value of $\alpha$'s purity~\cite{Piani2012}. 
However, if the ancilla does not participate in the dynamics then we have $\| \partial_t (\rho_t \otimes \alpha) \| = \| \partial_t \rho_t \| \cdot \| \alpha \|$ and the denominator is affected by the same factor. Thus, we have $T_D (\rho \otimes \alpha ,\sigma \otimes \alpha) =T_D (\rho,\sigma)$. However, if the ancillary system were to be correlated with the system or be part of the dynamics, then the actual time and the QSL will indeed be affected.

\subsection{Robustness under mixing}
\begin{figure}
    \centering
    \includegraphics[width=0.47\textwidth]{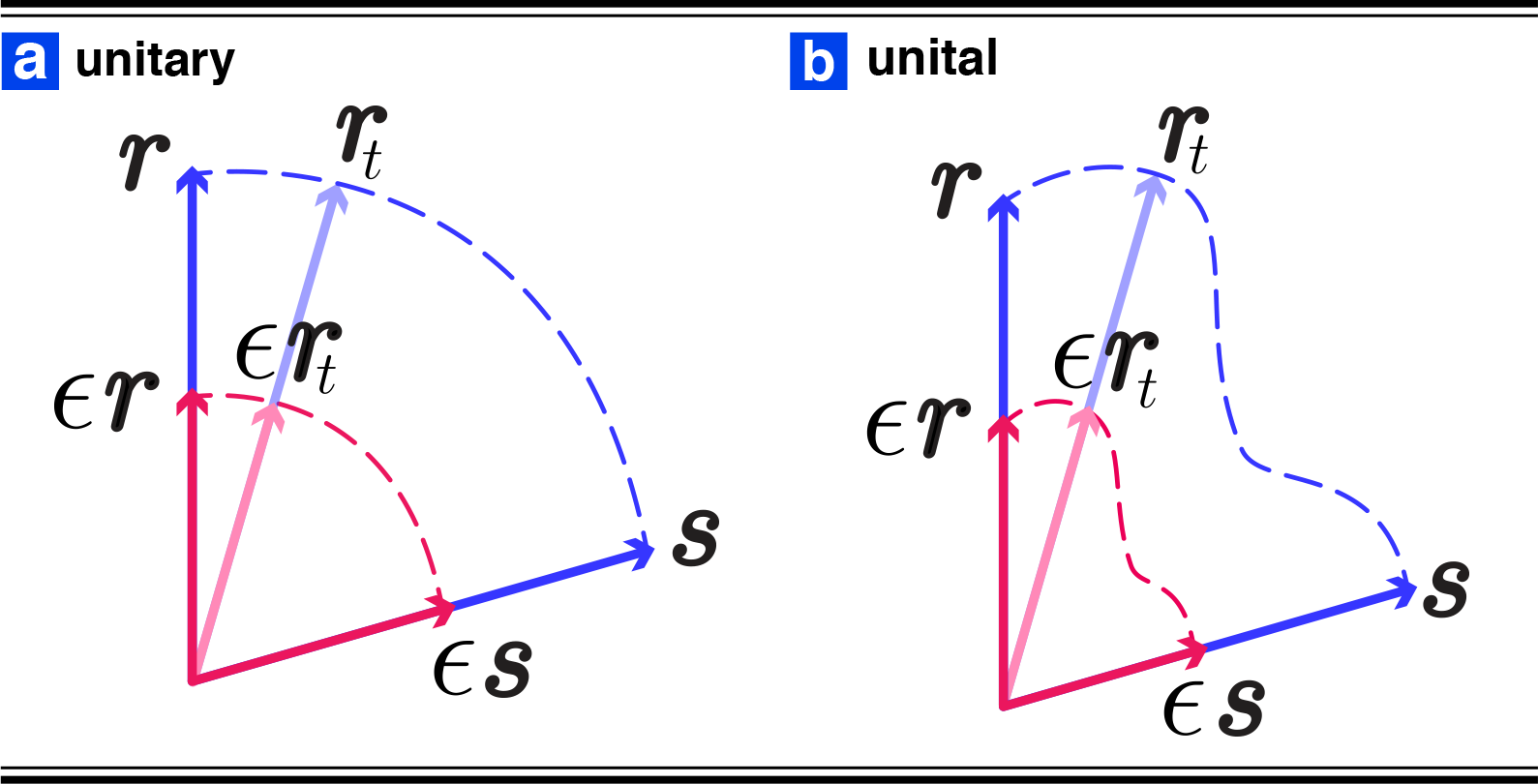} 
    \caption{\textbf{Robustness of $T_D$ under pure depolarization.} Bound $T_D$, expressed in Eq.~\eqref{eq:speed_limit_arbitrary}, is invariant under the effect of a pure depolarization channel $D_\epsilon:\rho\to\rho' =\epsilon\rho+\frac{1-\epsilon}{d}\mathbb{1}$, applied to initial and final state, as long as the dynamics satisfies $\dot{\rho}'_t=\epsilon\dot{\rho}_t$. In this figure $\bm{r}$ and $\bm{s}$ are the generalized Bloch vectors associated with some initial and final states $\rho$ and $\sigma$, respectively. Accordingly, $\epsilon\bm{r}=\bm{r}'$ and $\epsilon\bm{s}=\bm{s}'$ represent the initial and final states under the effect of the pure depolarization channel $D_\epsilon$, while $\bm{r}_t$ and $\epsilon\bm{r}_t$ represent the evolved states $\rho_t$ and $\rho'_t$, respectively. In terms of the GBV, the condition for robustness can be expressed as $\dot{\bm{r}}'_t=\epsilon\dot{\bm{r}}_t$. ({\fontfamily{phv}\selectfont \textbf{a}}) Unitary evolution trivially satisfies this condition, since it preserves the length of the GBV. ({\fontfamily{phv}\selectfont \textbf{b}}) More generally, all unital maps, i.e., maps that preserve the identity, also satisfy the condition for invariance. In terms of the GBV, the condition for invariance simply reduces to $\bm{r}_t'=\epsilon\bm{r}_t$. Similar geometric arguments can be made if we consider completely positive maps with an arbitrary fixed point $\phi$, where the robustness is guaranteed under mixing with $\phi$.
    }
    \label{fig:invariance}
\end{figure}

The usual QSL is tight for unitary dynamics of pure states~\cite{Deffner2013}. However, it becomes rather loose for mixed states. The reason for this, as we show in detail in Ref.~\cite{Campaioli2018}, is that the Bures distance, given in Eq.~\eqref{eq:Bures}, monotonically decreases under mixing. That is, $B (\rho,\sigma) \geq B (\rho',\sigma')$, where 
\begin{gather}
\label{eq:depolarization}
    \rho'=\mathcal{D}_\epsilon[\rho]:=\epsilon \rho + \frac{1-\epsilon}{d}\mathbb{1},
\end{gather}
with $\mathbb{1}$ being the identity operator and $\epsilon\in[0,1]$. When $\epsilon$ tends to 0, the Bures distance vanishes faster than the speed term, and so does the corresponding QSL. Now, note that the GBVs of $\rho$ ($\sigma$) and $\rho'$ ($\sigma'$) are $\bm{r}$ ($\bm{s}$) and $\epsilon\bm{r}$ ($\epsilon\bm{s}$) respectively. A unitary transformation that maps $\bm{r}$ to $\bm{s}$ will also map $\epsilon\bm{r}$ to $\epsilon\bm{s}$ in exactly the same time. That is, the value of $\epsilon$ is inconsequential. Based on this observation, we proposed the angle between the GBVs as distance because it is independent of $\epsilon$ and therefore robust under mixing, (see Fig.~\ref{fig:invariance} {\fontfamily{phv}\selectfont \textbf{a}}). This robustness is precisely the reason for the supremacy of the QSL provided in Ref.~\cite{Campaioli2018} over the usual QSL.

Even when open evolution is considered, it is of primary importance for a QSL to remain effective and tight for increasingly mixed initial and final states. We now show that the bound $T_D$, introduced in Eq.~\eqref{eq:speed_limit_arbitrary}, is robust under mixing not only for arbitrary unitary dynamics, but also for any open evolution with a well defined fixed point.

Our present bound is invariant when the initial state is mixed with the fixed state of the dynamics. Let the dynamics from $\rho$ to $\sigma$ be due to a completely positive linear map $\mathcal{C}$ with a fixed point $\phi$, i.e., $\mathcal{C}(\rho)=\sigma$ and $\mathcal{C}(\phi)=\phi$. If we change the initial state $\rho$ to $\rho'=\epsilon \rho + (1-\epsilon) \phi$, the the final state will be $\sigma'=\epsilon \sigma + (1-\epsilon) \phi$. This shrinks the numerator of the Eq.~\eqref{eq:speed_limit_arbitrary} by $\epsilon \in [0,1]$, i.e., $\|\rho'-\sigma'\| = \epsilon \|\rho-\sigma\|$. However, the denominator also shrinks by the same amount. If the dynamics is time independent then we also have $\dot{\phi}=0$ and $\dot{\rho}$ will be mapped to $\dot{\rho'}=\epsilon \dot{\rho}$.
Hence $T_D(\rho',\sigma')=T_D(\rho,\sigma)$. 
A similar, but more elaborate, argument can also be carried out for time dependent dynamics, but will be omitted from the present manuscript. The above result contains the previous case of unitary dynamics, and all unital dynamics, as they preserve the maximally mixed state. In this case the condition for robustness under mixing simply becomes a condition on the contraction factor for the length of the GBV $\bm{r}'_t=\epsilon \bm{r}_t$, as expressed in Fig.~\ref{fig:invariance} {\fontfamily{phv}\selectfont \textbf{b}}.

In the next section we study the form of the bound, with particular attention to the speed, for the fundamental types of quantum evolution.

\section{Form of the speed}
\label{s:forms}

\subsection{Unitary evolution} 
When \emph{unitary evolution} is considered, the denominator of Eq.~\eqref{eq:speed_limit_arbitrary} is a simple function of the time-dependent Hamiltonian $H_t$
\begin{gather}
    \label{eq:unitary_denominator}
     \overline{\lVert \dot{\rho}_t \rVert}  = \overline{\sqrt{2\;\tr[H_t^2\rho_t^2-(H_t\rho_t)^2]}}.
\end{gather}
This term is proportional (up to a constant of motion) to the term in the denominator of the QSL derived in Ref.~\cite{Campaioli2018}~\footnote{For pure states, this quantity reduces to the  time-averaged standard deviation of $H_t$, $\Delta E = \overline {\sqrt{\tr[H_t^2 \rho_t] - |\tr[H_t\rho_t]|^2}}$.}. However, the numerator of the QSL in Eq.~\eqref{eq:speed_limit_arbitrary} and the QSL in Ref.~\cite{Campaioli2018} are different and the latter is always tighter. This is because, for this special case, the length of the  GBV $\bm{r}$ must be preserved along the evolution~\cite{Campaioli2018}, and the geodesics are arcs of great circles that connect $\bm{r}$ and $\bm{s}$ (see Fig.~\ref{fig:distances}{ \fontfamily{phv} \selectfont \textbf{a}}). Accordingly, as we showed in Ref.~\cite{Campaioli2018}, the distance becomes $D_U(\rho,\sigma)=\Theta \lVert \bm{r}\rVert_2 = \Theta \lVert \bm{s}\rVert_2$, where
\begin{gather}
    \label{eq:generalized_bloch_angle}
    \Theta(\rho,\sigma) = \arccos(\hat{\bm{r}}\cdot\hat{\bm{s}}),
\end{gather}
is the generalized Bloch angle, with $\hat{\bm{r}}$, $\hat{\bm{s}}$ being the unit vectors associated to $\bm{r}$ and $\bm{s}$, respectively~\footnote{The distance defined in Ref.~\cite{Campaioli2018} differs by a constant factor $\lVert \bm{r}\rVert_2=\lVert \bm{s}\rVert_2$, which does not affect the derived QSL in the unitary case.}.

The above observation should not be surprising. The distance in the last equation corresponds to the arc-length, which is always greater than the distance in Eq.~\eqref{eq:speed_limit_arbitrary} measuring the length along the straight line. If we are promised that the evolution is unitary, then we are free to work with the tighter QSL from Ref.~\cite{Campaioli2018}. However, if that information is not available, we must be conservative and work with the QSL in Eq.~\eqref{eq:speed_limit_arbitrary}.

We now consider the open evolution case, starting with Lindblad dynamics, before proceeding with more general non-Markovian evolutions.

\subsection{Lindblad dynamics} 
In the case of semigroup dynamics, $ \overline{\lVert \dot{\rho}_t \rVert} $ becomes a function of the Lindblad operators~\cite{DelCampo2013}. As this function is generally complicated, we present its form for some particular types of Lindblad dynamics, for which it substantially simplifies. Let us consider a general form of the Lindblad master equation given by $\dot{\rho}_t=-i[H,\rho_t]+\sum_k\gamma_k\big(L_k\rho L_k^\dagger -\frac{1}{2}\{L_k^\dagger L_k,\rho_t\}\big)$, where typically the Lindblad operators are chosen to be orthonormal and traceless, i.e., $\tr[L_k L_l]=\delta_{kl}$, and $\tr[L_k]=0$. If the unitary part of the dynamics is irrelevant with respect to the dissipator, i.e., when $[H,\rho_t]$ is negligible when compared to the other terms, we obtain
\begin{align}
    \label{eq:lindblad_depolarization_denominator}
    \overline{\lVert \dot{\rho}_t \rVert} \leq 2 \sum_k \gamma_k^2 \overline{\lVert L_k \rVert^2},
\end{align}
where the inequality holds since $\lVert\rho_t\rVert \lVert\dot{\rho}_t\rVert \leq 2\sum_k \gamma_k^2 \lVert L_k\rVert^2 \lVert\rho_t\rVert ^2$~\cite{Uzdin2016}, and $\lVert \rho_t\rVert \leq 1$. We can readily apply this result to three important cases: pure dephasing dynamics, pure depolarization dynamics, and speed of purity change.\vspace{5pt} 

\textbf{Pure dephasing dynamics. ---} This type of dynamics models the idealized evolution of an open quantum system whose coherence decays over time due to the interaction with the environment. Under this kind of dynamics, a quantum system that evolves for a sufficiently long time is expected to lose its quantum mechanical features and exhibit a classical behavior.
Here, for the sake of clarity, we consider the case of \emph{pure dephasing of a single qubit}, described by the Lindblad equation $\dot{\rho}_t = \gamma(\sigma_z\rho_t\sigma_z -\rho_t)$, as done in Ref.~\cite{DelCampo2013}. The instantaneous speed reads $\lVert \dot{\rho}_t\rVert =\sqrt{2} \;\gamma \sqrt{r_1^2(t)+r_2^2(t)}$, where $\bm{r}_t = (r_1(t),r_2(t),r_3(t))$ is the Bloch vector associated to $\rho_t$. In this case the time-averaged speed can be bounded as
\begin{gather}
    \label{eq:dephasing_denominator}
    \overline{\lVert \dot{\rho}_t \rVert}  \leq \sqrt{2}\gamma.
\end{gather}
Although considering the case of a single qubit might sound simplistic, the same description can be used to cover relevant high-dimensional systems that effectively behave like qubits~\cite{Ilichev2003,Zueco2009}.\vspace{5pt} 

\textbf{Pure depolarizing dynamics. ---} Another interesting observation is that our bound $T_D$ is geometrically tight when purely depolarizing dynamics is considered, i.e., when $\rho_t= \mathcal{D}_{\epsilon(t)}[\rho_0] = \epsilon(t)\rho_0+\frac{1-\epsilon(t)}{d}\mathbb{1}$, which serves as an idealized model of noise for the evolution of an open quantum system that monotonically deteriorates towards the state of maximal entropy, i.e., the maximally mixed state. Geometrically, it corresponds to the re-scaling of the generalized Bloch vector $\bm{r}_t = \epsilon(t)\; \bm{r}_0$, where $\epsilon(0)=1$. Tightness is guaranteed by the fact that each vector $\bm{r}_t$ obtained in this way represents a state, along with the fact that the orbit of such evolution is given by the straight line that connects $\bm{r}_0$ to $\bm{r}_\tau$, whose length is exactly given by $D(\rho_0,\rho_\tau)$. In this case our bound reads
\begin{gather}
    T_D(\rho_0,\rho_\tau)=\frac{1-\epsilon(\tau)}{\overline{|\dot{\epsilon}(t)|}}.
\end{gather}
If we restrict ourselves to the case of strictly monotonic contraction (expansion) of the GBV, the denominator becomes $(1-\epsilon(\tau))/\tau$, which further supports our argument for the tightness of our bound. In this case, it simply returns the condition for optimal evolution $T_D(\rho_0,\rho_\tau)=\tau$, i.e., the evolution time $\tau$ coincides with the bound, and thus with the minimal time.\vspace{5pt} 

\textbf{Speed of purity change. ---} Since a contraction of the GBV corresponds to a decrease of the purity of the initial state $\rho_0$, Eq.~\eqref{eq:speed_limit_arbitrary} provides a QSL for the variation in the purity $\Delta \mathcal{P}$, which is saturated when obtained by means of purely depolarizing dynamics with strictly monotonic contraction. In particular, $\bm{r}\to\epsilon\bm{r}$ implies a variation of the purity $\tr[\rho_0^2]\to \epsilon^2\tr[\rho_0^2]+(1-\epsilon^2)\mathbb{1}/d$, which thus depends also on the dimension $d$ of the system. Similar QSLs have been derived by the authors of Ref.~\cite{Rodriguez-Rosario2011}, who express a bound on the instantaneous variation of the purity in terms of the strength of the interaction Hamiltonian and the properties of the total system-environment density operator, as well as by the authors of Ref.~\cite{Uzdin2016}, who provide a bound on the variation of the purity $\mathcal{P}[\rho_0]/\mathcal{P}[\rho_\tau]$ as a function of the non-unitary part of the evolution, both in Hilbert and Liouville space.

\subsection{Memory kernel master equation} 
For the most general non-Markovian dynamics, the denominator of bound~\eqref{eq:speed_limit_arbitrary} can be written in terms of a convolution with a memory kernel~\cite{Breuer2002}, e.g., in the form of the Nakajima-Zwanzig equation, $\dot{\rho}_t = \mathcal{L}_t\rho_t + \int_{t_0}^t ds\mathcal{K}_{t,s}\rho_s +\mathcal{J}_{t,t_0}$, where $\mathcal{L}_t$ is a time-local generator like that of the Lindblad master equation, the memory kernel $\mathcal{K}_{t,s}$ accounts for the effect of memory, and $\mathcal{J}_{t, t_0}$ accounts for initial correlations between system and environment~\cite{Pollock2017}. The denominator of bound~\eqref{eq:speed_limit_arbitrary} can be simplified using the triangle inequality $\lVert A + B + C\rVert \leq \lVert A\rVert  +\lVert B\rVert +\lVert C\rVert $, at the cost of its tightness. Similarly, the memory kernel can be divided up into a finite sum of terms whenever it is possible to resort to a temporal discretization, $\lVert\int_{t_0}^t ds\mathcal{K}_{t,s} \rho_s \rVert \sim\delta t\sum_k \lVert \mathcal{K}_{t_{k},t_{k-1}} \rho_{t_{k-1}} \rVert $, again, at the cost of reducing the tightness of the bound.

\textbf{Underlying evolution. ---} Alternatively, the orbit dependent term can always be related to an underlying unitary evolution with a wider environment: $\dot{\rho}_t = -i \ \tr_E [H,\Pi_t]$, where $H$ and $\Pi_t$ are the Hamiltonian and the state of the joint system-environment, respectively. We can further break down the total Hamiltonian into $H = H_S+H_{\textrm{int}} +H_E $, where $H_S$ ($H_E$) is the Hamiltonian of the system (environment) and $H_{\textrm{int}}$ describes the interactions between the two. In this case the denominator of bound~\eqref{eq:speed_limit_arbitrary} reads
\begin{align}
    \label{eq:arbitrary_denominator}
    \overline{\lVert \dot{\rho}_t \rVert}  &= \overline{\lVert -i[H_S,\rho_t]-i\tr_E\{[H_{\textrm{int}}, \Pi_t]\}\rVert},
\end{align}
since $\tr_E\{[H_E, \Pi_t]\} = 0$. A less tight speed limit can be obtained by splitting the right hand side of Eq.~\eqref{eq:arbitrary_denominator}, using the triangle inequality and the linearity of the time average, to obtain $\overline{\lVert \dot{\rho}_t \rVert} \leq \overline{\lVert -i[H_S,\rho_t]\rVert}  + \overline{\lVert-i\tr_E\{[H_{\textrm{int}}, \Pi_t]\}\rVert}$, in order to isolate the contribution of $H_{\textrm{int}}$ from that of $H_S$, when convenient. 

Additionally, by considering the larger Hilbert space of system and environment combined, it is possible to appreciate the difference between the traditional QSL, $T_B(\rho,\sigma) = B(\rho,\sigma) / \overline{\Delta E}$, expressed in terms of the Bures angle $B(\rho,\sigma)$ (see Eq.~\eqref{eq:Bures}), and the bound $T_D$ of Eq.~\eqref{eq:speed_limit_arbitrary}. The Bures distance $B(\rho,\sigma)$ corresponds to the minimal Fubini-Study distance between purifications of $\rho$ and $\sigma$ in a larger Hilbert space \cite{Bengtsson2008}, here denoted by $\ketbra{\psi}{\psi}$ and $\ketbra{\varphi}{\varphi}$, respectively. Such purified states must be entangled states of system and environment when $\rho$ and $\sigma$ are mixed. Moreover, unlike in Eq.~\eqref{eq:arbitrary_denominator}, these states may have nothing to do with the actual system-environment evolution. In general, in order to saturate the traditional QSL, one must have access to those (possibly fictional) entangled states, and be able to perform highly non-trivial operations over both system and environment, such as $\ketbra{\psi}{\varphi}+\ketbra{\varphi}{\psi}$, which can contain terms with high order of interaction \cite{Campaioli2017, Campaioli2018}. Since in practice one has little, if any control over the environment degrees of freedom, and nearly no access to the entangled state of the system and environment combined, the traditional QSL rapidly loses its efficacy.

In contrast, bound $T_D$, introduced in Eq.~\eqref{eq:speed_limit_arbitrary}, provides a conservative estimate of the minimal evolution time between two states $\rho$ and $\sigma$, under the assumption of no access to their purification. The speed of the evolution is assessed observing only the local part (the system) of a global evolution (the underlying unitary evolution of system and environment), as expressed by Eq.~\eqref{eq:arbitrary_denominator}, while still allowing for optimal driving of the purifications of $\rho$ and $\sigma$.
In addition to the ability of QSLs to represent an achievable bound for the minimal evolution time, their usefulness also depends on how easily they can be calculated and measured. We discuss this aspect in the next section, comparing the features of our bound $T_D$ to those of other QSLs.

\section{Feasibility}
As discussed in the introduction, the usefulness of a QSL bound is directly linked to the feasibility of its evaluation, whether it be computational or experimental. There are two types of difficulties that one might encounter in the evaluation of a QSL. First, computing the distance, that in our case is given by the orbit-independent term in the numerator of Eq.~\eqref{eq:speed_limit_arbitrary}, and  second, evaluating the speed, that in our case is given by the orbit-dependent term in the denominator of Eq.~\eqref{eq:speed_limit_arbitrary}. While the latter is usually related to some norm of $\dot{\rho}_t$, the former changes remarkably from bound to bound. We address the distance first, before proceeding to a discussion of the speed.\vspace{5pt} 

\textbf{The distance. ---} Among all the QSL bounds known so far one can make a clear-cut distinction between the type of distances that have been used: either they require evaluating the overlap $\tr[\rho\sigma]$ between the initial and the final states~\cite{Sun2015, DelCampo2013, Campaioli2018}, or they require to calculate $\sqrt{\rho}$ and $\sqrt{\sigma}$ (or similar functions)~\cite{Deffner2013b, Mondal2016, Pires2016}. The latter is much more complicated than the former, as it requires solving the eigenvalue and eigenvectors of $\rho$ and $\sigma$. While solving the former does not require diagonalizing the density matrices. Moreover, the overlap between two density operators ($\tr[\rho \sigma]$) is easily measured experimentally using a controlled \textsc{SWAP} and measurement on an ancillary system~\cite{Ekert2002} independent of the dimensions of the system.

The same approach can be used to estimate the fidelity between $\rho$ and $\sigma$, i.e., $\tr[\sqrt{\sqrt{\rho}\:\sigma\sqrt{\rho}}]$, but the number of interference experiments required grows with the dimension of the system in order to reach a good approximation. Conveniently, our bound $T_D$ simply depends on the overlap $\tr[\rho\sigma]$: This feature elevates $T_D$ to the most favourable choice, even in the cases where its tightness is comparable to that of other QSLs.\vspace{5pt} 

\textbf{The speed. ---}
We now move on to the orbit-dependent term  $\overline{\lVert \dot{\rho}_t\rVert}$, i.e., the denominator of Eq.~\eqref{eq:speed_limit_arbitrary}, which appears in different forms in virtually every QSL bound. This term can be interpreted as the \textit{speed} of the evolution~\footnote{Note that $\dot{\rho}_t$ is proportional to the tangent vector $\dot{\bm{r}}_t$, which can be regarded to as the velocity. Accordingly, the norm of the latter if the speed, and it is proportional up to a constant of motion to the HS norm of $\dot{\rho}_t$.}, and it can be hard to compute, as it might require the knowledge of the solution $\rho_t$ to the dynamical problem $\dot{\rho}_t=L[\rho_t]$. For this reason, one might criticize QSLs as being impractical, or ineffective, if too hard to compute. Surely, when QSLs are easy to compute, they can be used to quickly estimate the evolution time $\tau$, required by some specific dynamics $\dot{\rho}_t=L[\rho_t]$ to evolve between $\rho$ and $\sigma$, however, their main purpose is rather to answer the question, \emph{can we evolve faster?} The evaluation of a QSL bound for the initial and final states $\rho$ and $\sigma$, along the orbit described by $\rho_t$, immediately tells us if we could evolve faster along another orbit that has the same speed, or confirms that we are already doing the best we can.

Besides, it is not always necessary to solve the the dynamics of the system in order to evaluate the speed, which can be constant along the orbit. For example, the standard deviation of any time-independent Hamiltonian $H$ is a constant of motion, and can be directly obtained from the initial state of the system and the Hamiltonian $H$, making the speed extremely easy to compute. 
In the more general case of an actually orbit-dependent speed, it is often possible to numerically and experimentally estimate  $\overline{\lVert \dot{\rho}_t\rVert}$ using the following approach: First, we can approximate $\dot{\rho}_t$ with the finite-time increment, $\dot{\rho}_t \sim (\rho_{t_2}-\rho_{t_1})/|t_2-t_1|$, where $t_{1,2} = t \pm \epsilon/2$, for small $\epsilon$.
We then proceed with the approximation 
\begin{gather}
    \label{eq:finite_time}
    \tr[\dot{\rho}_t^2]  \sim \frac{\tr[\rho_{t_2}^2]+\tr[\rho_{t_1}^2] - 2\tr[\rho_{t_2}\rho_{t_1}]}{|t_2-t_1|^2}.
\end{gather}
Each term on the numerator of the right-hand side of Eq.~\eqref{eq:finite_time} can be evaluated with a controlled-\textsc{SWAP} circuit, as one would do for $\tr[\rho\sigma]$, as described above and in Ref.~\cite{Ekert2002}. 
In this sense, the Euclidean metric considered here has an advantage over those featuring $\sqrt{\rho_t}$, such as those based on quantum fidelity and affinity, since in general it requires fewer measurements for each instantaneous sample of the speed of the evolution.
Nevertheless, obtaining a precise estimation of the average speed of the evolution is generally hard, requiring a number of samples that strongly depends on the distribution of the velocities of the considered process, independently of the notion of the considered metric. When such estimation has to be approached, it is thus fundamental to reduce the amount of measurements required to obtain each instantaneous sample of the speed of the evolution.
In the next section, we will show that, in addition to being more feasible, our bound also outperforms existing speed limits for the majority of processes.

\section{Tightness}
\label{s:tightness}
\emph{Tightness of QSL bounds.}---
As stated in the introduction, one of our main interests is the performance of our bound $T_D$, in particular its tightness relative to other proposed QSLs. To this end, we must ensure that different bounds are fairly compared: Since QSL bounds depend on the orbit, they can only be compared with each other when evaluated along a chosen evolution, given fixed initial and final states. If their orbit-dependent terms are identical for any given evolution, such a comparison reduces to the evaluation of their orbit-independent terms. We compare our bound to the most significant bounds appearing in the literature~\cite{Sun2015, DelCampo2013, Deffner2013b, Pires2016,Mondal2016} which either depend on the overlap $\tr[\rho\sigma]$ or require the evaluation of quantum fidelity $F(\rho,\sigma) = \tr[\sqrt{\sqrt{\rho}\sigma\sqrt{\rho}}]$~\cite{Uhlmann1992b}, affinity $A(\rho,\sigma)=\tr[\sqrt{\rho}\sqrt{\sigma}]$~\cite{Luo2004}, or Fisher information~\cite{Facchi2010}.

We begin by considering the bound of Pires et al.~\cite{Pires2016}. In fact, Ref.~\cite{Pires2016} gives an infinite family of bounds that can be adjusted to fit the particular type of evolution that one might consider. However, as we pointed out in Ref.~\cite{Campaioli2018}, this freedom of choice has a drawback: The task of finding the right distance that induces a tight bound for the desired evolution is a difficult one. In addition, these bounds require the calculation of quantum fidelity or quantum Fisher information, which are almost always harder to evaluate and measure in comparison with the Hilbert-Schmidt norm (as discussed above). For these reasons, we disregard the family of bounds in Ref.~\cite{Pires2016}  from our subsequent discussion and proceed with the more practically feasible ones.

As different bounds can be meaningfully compared only when evaluated along the same orbit, one might be led to assume that the hierarchy between the bounds depends on the process in question. However, the orbit-dependent term that appears at the denominator of  bounds from Refs.~\cite{DelCampo2013,Deffner2013b,Sun2015} is always given by $\overline{\lVert \dot{\rho}_t\rVert}$~\footnote{In particular, we selected the Hilbert-Schmidt norm for analytical comparison, while we have  evaluated operator norm and trace numerically, if required by the considered QSL.} (i.e., the \emph{strength} of the generator), or can be directly related to it, up to some orbit-independent factors. This fact allows us to reduce the hierarchy of the bounds to that of the distance terms that depend only on the initial and final states, regardless of the chosen process and orbit. When this direct comparison is not possible, such as for the case of the bound in Ref.~\cite{Mondal2016}, we need to resort to numerical comparison.

The orbit-independent term of our bound can be directly compared with those of Sun et al.~\cite{Sun2015} and Del Campo et al.~\cite{DelCampo2013}, which depend on the overlap $\tr[\rho\sigma]$. In order to analytically compare our bound to that of Deffner et al.~\cite{Deffner2013b}, we over-estimate the orbit-independent term of the latter by replacing the fidelity with its lower-bound sub-fidelity, introduced in~\cite{Miszczak2008}, which  depends on the overlap $\tr[\rho\sigma]$, and on the additional quantity $\tr[(\rho\sigma)^2]$. For brevity, we will henceforth refer to previously introduced bounds by the corresponding first author's name.

\begin{figure}[b!]
    \centering
    \includegraphics[width=0.48\textwidth]{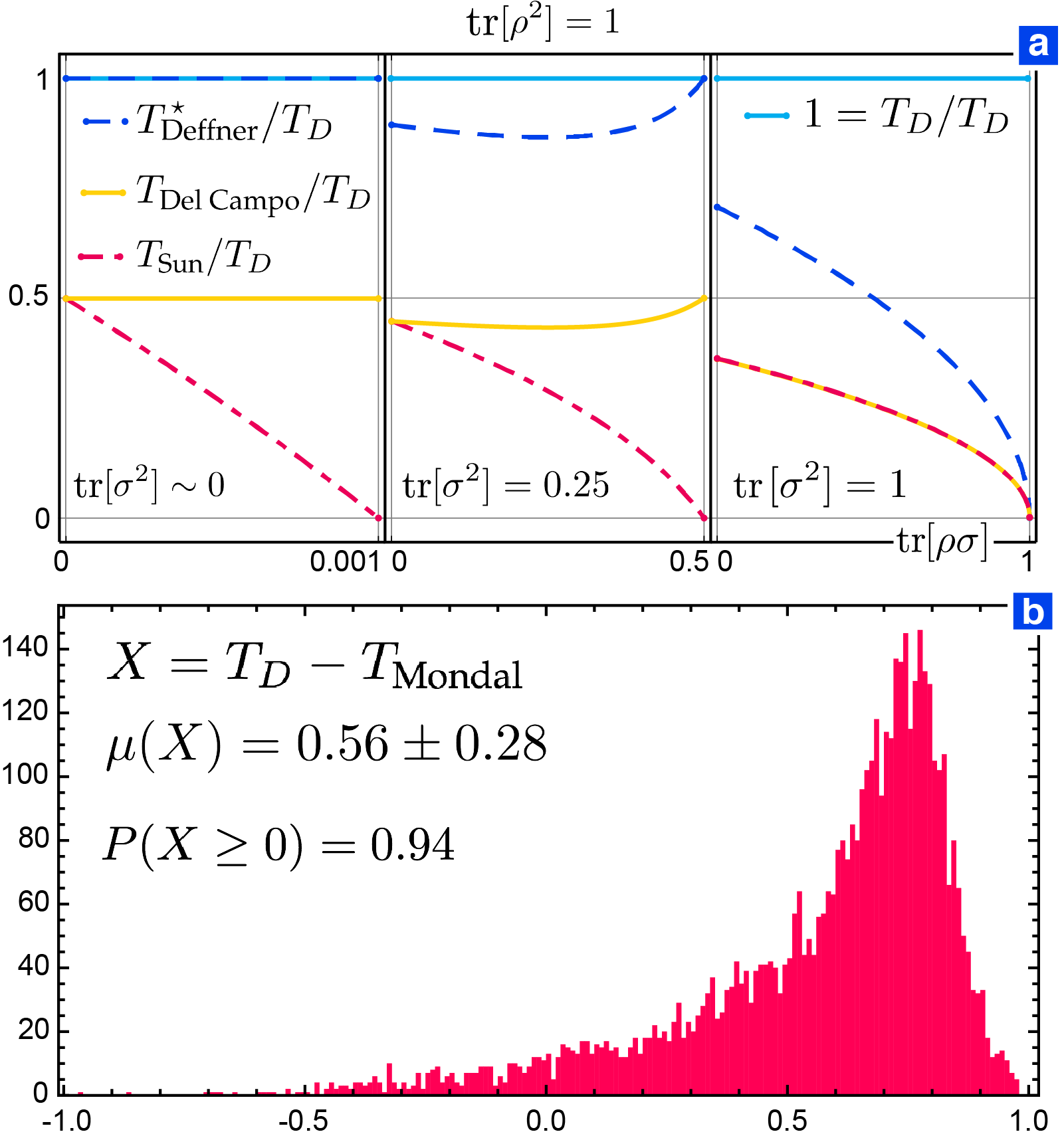}
    \caption{\textbf{Relative tightness of QSL bounds.}
    ({\fontfamily{phv}\selectfont \textbf{a}}) Analytic comparison of bounds from Refs.~\cite{Deffner2013b,DelCampo2013,Sun2015} with $T_D$, introduced in Eq.~\eqref{eq:speed_limit_arbitrary}, for the arbitrary evolution between $\rho=\rho^2$, and $\sigma$, expressed as the ratio between the considered QSL $T_{\textrm{Author}}$ and  $T_D$ as a function of $\tr[\rho\sigma]$ (which for $\rho^2=\rho$ also determines $\tr[(\rho\sigma)^2]$), as specified in the legend, and where $T_{\textrm{Deffner}}^\star\geq T_{\textrm{Deffner}}$ (see Eq.~\eqref{eq:over_deffner} in Appendix \ref{a:tightness} for details). The three insets represent three different choices of purity $\tr[\sigma^2]$ for the final state, from left to right $\tr[\sigma^2]\sim 0$, $\tr[\sigma^2]=0.25$,  $\tr[\sigma^2]=1$. Note that, since the dimension $d$ of the system sets a bound for the minimal value $1/d$ of $\tr[\rho^2]$, the central inset is meaningful from $d\geq 4$, while the left one is to be used in the limit of large $d$.
    ({\fontfamily{phv}\selectfont \textbf{b}}) Numerical estimation of relative tightness between $T_D$ and $T_{\textrm{Mondal}}$, obtained sampling $5000$ orbits from Haar-randomly distributed $H$, $\gamma_E$, and $\rho_0$, as described in section~\ref{s:tightness}. The dimension of the system's and environment's Hilbert spaces is uniformly sampled between $2$ and $10$. The tightness is measured using the parameter $X=T_D-T_{\textrm{Mondal}}$, which is bounded between $-1$ and $1$, given that the evolution is carried for a unit time $\tau=1$, and that both bounds have to be smaller than the evolution time $\tau$. The probability $P(X\geq 0)$ for our bound to be better than Mondal's is equal to $0.94$, with an average difference $\mu=0.56\pm0.28$.}
    \label{fig:hierarchy}
\end{figure}

As a result we find that, independently of the chosen process (i.e., for every choice of the generator $\dot{\rho}_t$), the bound $T_D$ expressed in Eq.~\eqref{eq:speed_limit_arbitrary} is tighter (i.e., greater) than Sun's, Del Campo's, and Deffner's for every (allowed) choice of $\rho$ and $\sigma$
\begin{align}
    \label{eq:sun_delcampo}
   & T_D \geq \max\big\{T_{\textrm{Sun}},T_{\textrm{Del Campo}}\big\}, \;\; \forall \rho,\sigma \; \in \mathcal{S}(\mathcal{H}_S), \\
   \label{eq:deffner_beat}
   & T_D \geq T_{\textrm{Deffner}} \;\; \forall \rho,\sigma, \; \textrm{s.t.} \; \rho^2=\rho  \; \textrm{or} \; \sigma^2=\sigma,
\end{align}
(see Fig.~\ref{fig:hierarchy} { \fontfamily{phv}\selectfont \textbf{a}}). While Sun's and Del Campo's QSLs are as easy to compute as our QSL given in Eq.~\eqref{eq:speed_limit_arbitrary}, they are also the loosest bounds. In contrast, Deffner bound's can be as tight as ours, but, since it requires the evaluation of $\sqrt{\rho}$ and $\sqrt{\sigma}$, it is less feasible.
In particular, Deffner's bound has been proven to be valid only when one of the two states is pure, i.e., for $\rho=\rho^2$ (or $\sigma=\sigma^2$)~\cite{Sun2015}. Under this condition our bound is always tighter than Deffner's. Additionally, we can analytically extend the validity of Deffner's bound to a larger class of cases by comparing it with our bound, and studying the region of the space of states for which $T_D$ is larger. All the details about the relative tightness of the considered bounds can be found in Appendix~\ref{a:tightness}.

Finally, we compare our bound to that of Ref.~\cite{Mondal2016} by Mondal et al., derived for the case of any general evolution, starting from the assumption that the initial state of the system $\rho_0$ is uncorrelated with that of the environment $\gamma_E$, i.e., $\Pi_0=\rho_0\otimes\gamma_E$. The orbit-independent term of their bound is a function of the affinity $A(\rho_0,\rho_\tau)=\tr[\sqrt{\rho_0}\sqrt{\rho_\tau}]$ between initial and final states of the system, $\rho_0$ and $\rho_\tau$, respectively~\cite{Mondal2016}, which, as mentioned earlier, is hard to calculate and to measure as it requires the diagonalization of both density operators. The orbit-dependent term of their bound is a function of the root of $\rho_0$ and of an \emph{effective} Hamiltonian $\widetilde{H_S} = \tr_E[H\; \mathbb{1}\otimes\gamma_E]$, where $H$ is the total system-environment Hamiltonian. This function is not equivalent to $\overline{\lVert \dot{\rho}_t \rVert} $ (not even up to an orbit-independent factor), so we must calculate the two bounds for any given choice of dynamics, i.e., for any choice of total system-environment Hamiltonian $H$ and of initial state of the environment $\gamma_E$. 

As such, we proceed with a numerical comparison of the two bounds. We randomly generate total Hamiltonians $H$, initial states of the environment $\gamma_E$, and initial states of the system $\rho_0$, in order to choose the final state of the system $\rho_\tau = \tr_E [U_\tau \rho_0\otimes\gamma_E U_\tau^\dagger]$, where $U_\tau = \exp[-i H \tau]$, fixing $\tau=1$ for reference. We then compute both bounds for each instance of $H$, $\gamma_E$, and $\rho_0$ and compare their performance by measuring the difference $T_D - T_{\textrm{Mondal}}$. Remembering that $\tau=1$, and that both bounds must be smaller than $\tau$, the difference $T_D - T_{\textrm{Mondal}}$ must be bounded by $-1$ and $1$. Our numerical results provide a convincing evidence of the performance of $T_D$ over $T_{\textrm{Mondal}}$, with the former being larger then the latter in $94\%$ of the cases for the considered sample, with an average difference of $0.57\pm0.28$ (see Fig.~\ref{fig:hierarchy} {\fontfamily{phv}\selectfont \textbf{b}} for the details about the numerical study). While Mondal's bound performs better than Deffner's, Sun's and Del Campo's, it is arguably less feasible than all of them, as it involves the evaluation of $\sqrt{\rho}$ and $\sqrt{\sigma}$ for both distance and speed terms.

We have now shown that bound $T_D$ of Eq.~\eqref{eq:speed_limit_arbitrary} is tighter than the QSLs by Del Campo et al.~\cite{DelCampo2013}, by Sun et al.~\cite{Sun2015}, and by Deffner et al.~\cite{Deffner2013b}, for all processes, while being just as easy to compute (if not easier). We have also provided numerical evidence of the superiority of our bound $T_D$ over the QSL by Mondal et al.~\cite{Mondal2016} for almost all processes (for over 90\% instances), while being more feasible.

\section{Conclusions}
In this Article, we have presented a geometric quantum speed limit for arbitrary open quantum evolutions, based on the natural embedding of the space of quantum states in a high-dimensional ball, where states are represented by generalized Bloch vectors. Our speed limit $T_D$ is induced by the Euclidean norm of the displacement vector $\bm{r}-\bm{s}$ between the two generalized Bloch vectors $\bm{r}$ and $\bm{s}$, associated with the initial and final states of the evolution. The measure of distinguishability that arises from this choice of distance corresponds to the Hilbert-Schmidt norm of the difference between initial and final states, $\rho$ and $\sigma$. The use of this norm has several benefits: It allows for the effective use of optimization techniques, such as convex optimization and semidefinite programming \cite{Abernethy2009}, it is easy to manipulate analytically and numerically, it has a straightforward geometric interpretation, and it is independent from the choice of the Lie algebra $\bm{\Lambda}$ of $SU(d)$ used to represent states as GBVs. The Hilbert-Schmidt norm is also widely used in experimental context, not only for quantum optimal control tasks, in order to impose finite energy bandwidth constraints on the control Hamiltonian \cite{Wang2015,Geng2016}.

We have considered the case of general open dynamics, in terms of a system-environment Hamiltonian and convolution with a memory kernel, as well as the special cases of unitary evolution and Lindblad dynamics. While the performance of many QSLs diminishes when increasingly mixed states are considered, our bound remains robust under mixing, as well as under composition. We have discussed the form of the our bound, with particular attention given to the speed of the evolution. We highlighted the differences between our bound and the traditional QSL, induced by the Bures distance, and shed light on the reasons for the poor performance of the latter. Comparatively speaking, our bound outperforms several bounds derived so far in the literature~\cite{Sun2015, DelCampo2013, Deffner2013b, Mondal2016} for the majority (if not all) processes. We have also address the physical interpretation of our bound, as well as that of similar QSLs, by providing a feasible experimental procedure that aims at the estimation of both the distance $D$ and the speed of the evolution $\overline{\lVert \dot{\rho}_t \rVert}$, while showing that our bound is easier to compute, as well as experimentally measure, than the other comparably tight bounds~\cite{Deffner2013b, Pires2016, Mondal2016}. These features indicate $T_D$ as the preferred choice of QSL. In particular, the versatility of this bound, as compared to that of Ref.~\cite{Campaioli2018}, allows it to be used for an much larger class of dynamics, which we have only just approached with our examples in Section~\ref{s:forms}; a reflection that will hopefully be inspiring for further studies.

The efficacy of the QSL derived from this distance suggests that the use of a real vector space equipped with Euclidean metric to represent the space of operators could also find application in the search for constructive approaches to time-optimal state preparation and gate design. This geometric picture might also offer solutions to some urgent outstanding problems, such as quantum optimal control in the presence of uncontrollable drift terms and constraints on tangent space, local quantum speed limits for multipartite evolution with restricted order of the interaction, and time-optimal unitary design for high-dimensional systems. The restrictions imposed by the constraints on the generators of evolution are known to dramatically change the the geodesic that connects two states, and thus the bound on the minimal time of evolutions, as discussed in Refs.~\cite{Arenz_2014,Lee2018}. There, the authors introduced methods to bound the speed of evolution depending on the form of uncontrollable drift terms, control complexity and size of the system, obtaining accurate results for the case of single qubits in~Ref.~\cite{Arenz_2017}. Combining such considerations with the geometric approach used here could simplify the task of improving quantum speed limits and optimal driving of controlled quantum systems, by exploiting constants of motions that are easier to represent in the generalised Bloch sphere picture.

Adapting this approach could find applications in other areas of quantum information, such as quantum metrology and quantum thermodynamics, where geodesic equations and geometric methods are routinely employed for the solution of optimization problems. While an attainable speed limit for arbitrary processes is yet to be found, its comprehension goes hand in hand with the understanding of the geometry of quantum states, as well as with the development of constructive techniques for time-optimal control.

\begin{acknowledgments}
\noindent 
We kindly acknowledge R. Uzdin, and V. Giovannetti for the fruitful discussion. KM is supported through Australian Research Council Future Fellowship FT160100073.
\end{acknowledgments}

\bibliographystyle{apsrev4-1_custom}
\bibliography{final_version}

\begin{thebibliography}{74}%
\makeatletter
\providecommand \@ifxundefined [1]{%
 \@ifx{#1\undefined}
}%
\providecommand \@ifnum [1]{%
 \ifnum #1\expandafter \@firstoftwo
 \else \expandafter \@secondoftwo
 \fi
}%
\providecommand \@ifx [1]{%
 \ifx #1\expandafter \@firstoftwo
 \else \expandafter \@secondoftwo
 \fi
}%
\providecommand \natexlab [1]{#1}%
\providecommand \enquote  [1]{``#1''}%
\providecommand \bibnamefont  [1]{#1}%
\providecommand \bibfnamefont [1]{#1}%
\providecommand \citenamefont [1]{#1}%
\providecommand \href@noop [0]{\@secondoftwo}%
\providecommand \href [0]{\begingroup \@sanitize@url \@href}%
\providecommand \@href[1]{\@@startlink{#1}\@@href}%
\providecommand \@@href[1]{\endgroup#1\@@endlink}%
\providecommand \@sanitize@url [0]{\catcode `\\12\catcode `\$12\catcode
  `\&12\catcode `\#12\catcode `\^12\catcode `\_12\catcode `\%12\relax}%
\providecommand \@@startlink[1]{}%
\providecommand \@@endlink[0]{}%
\providecommand \url  [0]{\begingroup\@sanitize@url \@url }%
\providecommand \@url [1]{\endgroup\@href {#1}{\urlprefix }}%
\providecommand \urlprefix  [0]{URL }%
\providecommand \Eprint [0]{\href }%
\providecommand \doibase [0]{http://dx.doi.org/}%
\providecommand \selectlanguage [0]{\@gobble}%
\providecommand \bibinfo  [0]{\@secondoftwo}%
\providecommand \bibfield  [0]{\@secondoftwo}%
\providecommand \translation [1]{[#1]}%
\providecommand \BibitemOpen [0]{}%
\providecommand \bibitemStop [0]{}%
\providecommand \bibitemNoStop [0]{.\EOS\space}%
\providecommand \EOS [0]{\spacefactor3000\relax}%
\providecommand \BibitemShut  [1]{\csname bibitem#1\endcsname}%
\let\auto@bib@innerbib\@empty
\bibitem [{\citenamefont {Mandelstam}\ and\ \citenamefont
  {Tamm}(1945)}]{Mandelstam1945}%
  \BibitemOpen
  \bibfield  {author} {\bibinfo {author} {\bibfnamefont {L.}~\bibnamefont
  {Mandelstam}}\ and\ \bibinfo {author} {\bibfnamefont {I.}~\bibnamefont
  {Tamm}},\ }\bibfield  {title} {\enquote {\bibinfo {title} {{The Uncertainty
  Relation Between Energy and Time in Non-relativistic Quantum Mechanics}},}\
  }in\ \href {\doibase 10.1007/978-3-642-74626-0_8} {\emph {\bibinfo
  {booktitle} {Sel. Pap.}}}\ (\bibinfo  {publisher} {Springer Berlin
  Heidelberg},\ \bibinfo {address} {Berlin, Heidelberg},\ \bibinfo {year}
  {1945})\ pp.\ \bibinfo {pages} {115--123}\BibitemShut {NoStop}%
\bibitem [{\citenamefont {Margolus}\ and\ \citenamefont
  {Levitin}(1998)}]{Margolus1998}%
  \BibitemOpen
  \bibfield  {author} {\bibinfo {author} {\bibfnamefont {N.}~\bibnamefont
  {Margolus}}\ and\ \bibinfo {author} {\bibfnamefont {L.~B.}\ \bibnamefont
  {Levitin}},\ }\bibfield  {title} {\enquote {\bibinfo {title} {{The maximum
  speed of dynamical evolution}},}\ }\href {\doibase
  10.1016/S0167-2789(98)00054-2} {\bibfield  {journal} {\bibinfo  {journal}
  {Phys. D Nonlinear Phenom.}\ }\textbf {\bibinfo {volume} {120}},\ \bibinfo
  {pages} {188} (\bibinfo {year} {1998})}\BibitemShut {NoStop}%
\bibitem [{\citenamefont {Deffner}\ and\ \citenamefont
  {Lutz}(2013{\natexlab{a}})}]{Deffner2013}%
  \BibitemOpen
  \bibfield  {author} {\bibinfo {author} {\bibfnamefont {S.}~\bibnamefont
  {Deffner}}\ and\ \bibinfo {author} {\bibfnamefont {E.}~\bibnamefont {Lutz}},\
  }\bibfield  {title} {\enquote {\bibinfo {title} {{Energy–time uncertainty
  relation for driven quantum systems}},}\ }\href {\doibase
  10.1088/1751-8113/46/33/335302} {\bibfield  {journal} {\bibinfo  {journal}
  {J. Phys. A Math. Theor.}\ }\textbf {\bibinfo {volume} {46}},\ \bibinfo
  {pages} {335302} (\bibinfo {year} {2013}{\natexlab{a}})}\BibitemShut
  {NoStop}%
\bibitem [{\citenamefont {Deffner}\ and\ \citenamefont
  {Campbell}(2017)}]{Deffner2017}%
  \BibitemOpen
  \bibfield  {author} {\bibinfo {author} {\bibfnamefont {S.}~\bibnamefont
  {Deffner}}\ and\ \bibinfo {author} {\bibfnamefont {S.}~\bibnamefont
  {Campbell}},\ }\bibfield  {title} {\enquote {\bibinfo {title} {{Quantum speed
  limits: from Heisenberg's uncertainty principle to optimal quantum
  control}},}\ }\href {\doibase 10.1088/1751-8121/aa86c6} {\bibfield  {journal}
  {\bibinfo  {journal} {J. Phys. A Math. Theor.}\ }\textbf {\bibinfo {volume}
  {50}},\ \bibinfo {pages} {453001} (\bibinfo {year} {2017})}\BibitemShut
  {NoStop}%
\bibitem [{\citenamefont {Lloyd}(2000)}]{Lloyd2000}%
  \BibitemOpen
  \bibfield  {author} {\bibinfo {author} {\bibfnamefont {S.}~\bibnamefont
  {Lloyd}},\ }\bibfield  {title} {\enquote {\bibinfo {title} {{Ultimate
  physical limits to computation}},}\ }\href {\doibase 10.1038/35023282}
  {\bibfield  {journal} {\bibinfo  {journal} {Nature}\ }\textbf {\bibinfo
  {volume} {406}},\ \bibinfo {pages} {1047} (\bibinfo {year}
  {2000})}\BibitemShut {NoStop}%
\bibitem [{\citenamefont {Giovannetti}\ \emph {et~al.}(2003)\citenamefont
  {Giovannetti}, \citenamefont {Lloyd},\ and\ \citenamefont
  {Maccone}}]{Giovannetti2003a}%
  \BibitemOpen
  \bibfield  {author} {\bibinfo {author} {\bibfnamefont {V.}~\bibnamefont
  {Giovannetti}}, \bibinfo {author} {\bibfnamefont {S.}~\bibnamefont {Lloyd}},
  \ and\ \bibinfo {author} {\bibfnamefont {L.}~\bibnamefont {Maccone}},\
  }\bibfield  {title} {\enquote {\bibinfo {title} {{Quantum limits to dynamical
  evolution}},}\ }\href {\doibase 10.1103/PhysRevA.67.052109} {\bibfield
  {journal} {\bibinfo  {journal} {Phys. Rev. A}\ }\textbf {\bibinfo {volume}
  {67}},\ \bibinfo {pages} {1} (\bibinfo {year} {2003})}\BibitemShut {NoStop}%
\bibitem [{\citenamefont {Alipour}\ \emph {et~al.}(2014)\citenamefont
  {Alipour}, \citenamefont {Mehboudi},\ and\ \citenamefont
  {Rezakhani}}]{Alipour2014}%
  \BibitemOpen
  \bibfield  {author} {\bibinfo {author} {\bibfnamefont {S.}~\bibnamefont
  {Alipour}}, \bibinfo {author} {\bibfnamefont {M.}~\bibnamefont {Mehboudi}}, \
  and\ \bibinfo {author} {\bibfnamefont {A.~T.}\ \bibnamefont {Rezakhani}},\
  }\bibfield  {title} {\enquote {\bibinfo {title} {{Quantum Metrology in Open
  Systems: Dissipative Cram{\'{e}}r-Rao Bound}},}\ }\href {\doibase
  10.1103/PhysRevLett.112.120405} {\bibfield  {journal} {\bibinfo  {journal}
  {Phys. Rev. Lett.}\ }\textbf {\bibinfo {volume} {112}},\ \bibinfo {pages}
  {120405} (\bibinfo {year} {2014})}\BibitemShut {NoStop}%
\bibitem [{\citenamefont {Giovannetti}\ \emph {et~al.}(2011)\citenamefont
  {Giovannetti}, \citenamefont {Lloyd},\ and\ \citenamefont
  {Maccone}}]{Giovannetti2011}%
  \BibitemOpen
  \bibfield  {author} {\bibinfo {author} {\bibfnamefont {V.}~\bibnamefont
  {Giovannetti}}, \bibinfo {author} {\bibfnamefont {S.}~\bibnamefont {Lloyd}},
  \ and\ \bibinfo {author} {\bibfnamefont {L.}~\bibnamefont {Maccone}},\
  }\bibfield  {title} {\enquote {\bibinfo {title} {{Advances in quantum
  metrology}},}\ }\href {\doibase 10.1038/nphoton.2011.35} {\bibfield
  {journal} {\bibinfo  {journal} {Nat. Photonics}\ }\textbf {\bibinfo {volume}
  {5}},\ \bibinfo {pages} {222} (\bibinfo {year} {2011})}\BibitemShut {NoStop}%
\bibitem [{\citenamefont {Chin}\ \emph {et~al.}(2012)\citenamefont {Chin},
  \citenamefont {Huelga},\ and\ \citenamefont {Plenio}}]{Chin2012}%
  \BibitemOpen
  \bibfield  {author} {\bibinfo {author} {\bibfnamefont {A.~W.}\ \bibnamefont
  {Chin}}, \bibinfo {author} {\bibfnamefont {S.~F.}\ \bibnamefont {Huelga}}, \
  and\ \bibinfo {author} {\bibfnamefont {M.~B.}\ \bibnamefont {Plenio}},\
  }\bibfield  {title} {\enquote {\bibinfo {title} {{Quantum Metrology in
  Non-Markovian Environments}},}\ }\href {\doibase
  10.1103/PhysRevLett.109.233601} {\bibfield  {journal} {\bibinfo  {journal}
  {Phys. Rev. Lett.}\ }\textbf {\bibinfo {volume} {109}},\ \bibinfo {pages}
  {233601} (\bibinfo {year} {2012})}\BibitemShut {NoStop}%
\bibitem [{\citenamefont {Demkowicz-Dobrza{\'{n}}ski}\ \emph
  {et~al.}(2012)\citenamefont {Demkowicz-Dobrza{\'{n}}ski}, \citenamefont
  {Ko{\l}ody{\'{n}}ski},\ and\ \citenamefont
  {Guţă}}]{Demkowicz-Dobrzanski2012}%
  \BibitemOpen
  \bibfield  {author} {\bibinfo {author} {\bibfnamefont {R.}~\bibnamefont
  {Demkowicz-Dobrza{\'{n}}ski}}, \bibinfo {author} {\bibfnamefont
  {J.}~\bibnamefont {Ko{\l}ody{\'{n}}ski}}, \ and\ \bibinfo {author}
  {\bibfnamefont {M.}~\bibnamefont {Guţă}},\ }\bibfield  {title} {\enquote
  {\bibinfo {title} {{The elusive Heisenberg limit in quantum-enhanced
  metrology}},}\ }\href {\doibase 10.1038/ncomms2067} {\bibfield  {journal}
  {\bibinfo  {journal} {Nat. Commun.}\ }\textbf {\bibinfo {volume} {3}},\
  \bibinfo {pages} {1063} (\bibinfo {year} {2012})}\BibitemShut {NoStop}%
\bibitem [{\citenamefont {Chenu}\ \emph {et~al.}(2017)\citenamefont {Chenu},
  \citenamefont {Beau}, \citenamefont {Cao},\ and\ \citenamefont {del
  Campo}}]{Chenu2017}%
  \BibitemOpen
  \bibfield  {author} {\bibinfo {author} {\bibfnamefont {A.}~\bibnamefont
  {Chenu}}, \bibinfo {author} {\bibfnamefont {M.}~\bibnamefont {Beau}},
  \bibinfo {author} {\bibfnamefont {J.}~\bibnamefont {Cao}}, \ and\ \bibinfo
  {author} {\bibfnamefont {A.}~\bibnamefont {del Campo}},\ }\bibfield  {title}
  {\enquote {\bibinfo {title} {{Quantum Simulation of Generic Many-Body Open
  System Dynamics Using Classical Noise}},}\ }\href {\doibase
  10.1103/PhysRevLett.118.140403} {\bibfield  {journal} {\bibinfo  {journal}
  {Phys. Rev. Lett.}\ }\textbf {\bibinfo {volume} {118}},\ \bibinfo {pages}
  {140403} (\bibinfo {year} {2017})}\BibitemShut {NoStop}%
\bibitem [{\citenamefont {Reich}\ \emph {et~al.}(2012)\citenamefont {Reich},
  \citenamefont {Ndong},\ and\ \citenamefont {Koch}}]{Reich2012}%
  \BibitemOpen
  \bibfield  {author} {\bibinfo {author} {\bibfnamefont {D.~M.}\ \bibnamefont
  {Reich}}, \bibinfo {author} {\bibfnamefont {M.}~\bibnamefont {Ndong}}, \ and\
  \bibinfo {author} {\bibfnamefont {C.~P.}\ \bibnamefont {Koch}},\ }\bibfield
  {title} {\enquote {\bibinfo {title} {{Monotonically convergent optimization
  in quantum control using Krotov's method}},}\ }\href {\doibase
  10.1063/1.3691827} {\bibfield  {journal} {\bibinfo  {journal} {J. Chem.
  Phys.}\ }\textbf {\bibinfo {volume} {136}},\ \bibinfo {pages} {104103}
  (\bibinfo {year} {2012})}\BibitemShut {NoStop}%
\bibitem [{\citenamefont {Caneva}\ \emph {et~al.}(2009)\citenamefont {Caneva},
  \citenamefont {Murphy}, \citenamefont {Calarco}, \citenamefont {Fazio},
  \citenamefont {Montangero}, \citenamefont {Giovannetti},\ and\ \citenamefont
  {Santoro}}]{Caneva2009}%
  \BibitemOpen
  \bibfield  {author} {\bibinfo {author} {\bibfnamefont {T.}~\bibnamefont
  {Caneva}}, \bibinfo {author} {\bibfnamefont {M.}~\bibnamefont {Murphy}},
  \bibinfo {author} {\bibfnamefont {T.}~\bibnamefont {Calarco}}, \bibinfo
  {author} {\bibfnamefont {R.}~\bibnamefont {Fazio}}, \bibinfo {author}
  {\bibfnamefont {S.}~\bibnamefont {Montangero}}, \bibinfo {author}
  {\bibfnamefont {V.}~\bibnamefont {Giovannetti}}, \ and\ \bibinfo {author}
  {\bibfnamefont {G.~E.}\ \bibnamefont {Santoro}},\ }\bibfield  {title}
  {\enquote {\bibinfo {title} {{Optimal control at the quantum speed limit}},}\
  }\href {\doibase 10.1103/PhysRevLett.103.240501} {\bibfield  {journal}
  {\bibinfo  {journal} {Phys. Rev. Lett.}\ }\textbf {\bibinfo {volume} {103}},\
  \bibinfo {pages} {240501} (\bibinfo {year} {2009})}\BibitemShut {NoStop}%
\bibitem [{\citenamefont {del Campo}\ \emph {et~al.}(2012)\citenamefont {del
  Campo}, \citenamefont {Rams},\ and\ \citenamefont {Zurek}}]{DelCampo2012}%
  \BibitemOpen
  \bibfield  {author} {\bibinfo {author} {\bibfnamefont {A.}~\bibnamefont {del
  Campo}}, \bibinfo {author} {\bibfnamefont {M.~M.}\ \bibnamefont {Rams}}, \
  and\ \bibinfo {author} {\bibfnamefont {W.~H.}\ \bibnamefont {Zurek}},\
  }\bibfield  {title} {\enquote {\bibinfo {title} {{Assisted Finite-Rate
  Adiabatic Passage Across a Quantum Critical Point: Exact Solution for the
  Quantum Ising Model}},}\ }\href {\doibase 10.1103/PhysRevLett.109.115703}
  {\bibfield  {journal} {\bibinfo  {journal} {Phys. Rev. Lett.}\ }\textbf
  {\bibinfo {volume} {109}},\ \bibinfo {pages} {115703} (\bibinfo {year}
  {2012})}\BibitemShut {NoStop}%
\bibitem [{\citenamefont {Hegerfeldt}(2013)}]{Hegerfeldt2013}%
  \BibitemOpen
  \bibfield  {author} {\bibinfo {author} {\bibfnamefont {G.~C.}\ \bibnamefont
  {Hegerfeldt}},\ }\bibfield  {title} {\enquote {\bibinfo {title} {{Driving at
  the Quantum Speed Limit: Optimal Control of a Two-Level System}},}\ }\href
  {\doibase 10.1103/PhysRevLett.111.260501} {\bibfield  {journal} {\bibinfo
  {journal} {Phys. Rev. Lett.}\ }\textbf {\bibinfo {volume} {111}},\ \bibinfo
  {pages} {260501} (\bibinfo {year} {2013})}\BibitemShut {NoStop}%
\bibitem [{\citenamefont {Murphy}\ \emph {et~al.}(2010)\citenamefont {Murphy},
  \citenamefont {Montangero}, \citenamefont {Giovannetti},\ and\ \citenamefont
  {Calarco}}]{Murphy2010}%
  \BibitemOpen
  \bibfield  {author} {\bibinfo {author} {\bibfnamefont {M.}~\bibnamefont
  {Murphy}}, \bibinfo {author} {\bibfnamefont {S.}~\bibnamefont {Montangero}},
  \bibinfo {author} {\bibfnamefont {V.}~\bibnamefont {Giovannetti}}, \ and\
  \bibinfo {author} {\bibfnamefont {T.}~\bibnamefont {Calarco}},\ }\bibfield
  {title} {\enquote {\bibinfo {title} {{Communication at the quantum speed
  limit along a spin chain}},}\ }\href
  {https://doi.org/10.1103/PhysRevA.82.022318} {\bibfield  {journal} {\bibinfo
  {journal} {Phys. Rev. A}\ }\textbf {\bibinfo {volume} {82}},\ \bibinfo
  {pages} {022318} (\bibinfo {year} {2010})}\BibitemShut {NoStop}%
\bibitem [{\citenamefont {An}\ \emph {et~al.}(2016)\citenamefont {An},
  \citenamefont {Lv}, \citenamefont {del Campo},\ and\ \citenamefont
  {Kim}}]{An2016}%
  \BibitemOpen
  \bibfield  {author} {\bibinfo {author} {\bibfnamefont {S.}~\bibnamefont
  {An}}, \bibinfo {author} {\bibfnamefont {D.}~\bibnamefont {Lv}}, \bibinfo
  {author} {\bibfnamefont {A.}~\bibnamefont {del Campo}}, \ and\ \bibinfo
  {author} {\bibfnamefont {K.}~\bibnamefont {Kim}},\ }\bibfield  {title}
  {\enquote {\bibinfo {title} {{Shortcuts to adiabaticity by counterdiabatic
  driving for trapped-ion displacement in phase space}},}\ }\href {\doibase
  10.1038/ncomms12999} {\bibfield  {journal} {\bibinfo  {journal} {Nat.
  Commun.}\ }\textbf {\bibinfo {volume} {7}},\ \bibinfo {pages} {12999}
  (\bibinfo {year} {2016})}\BibitemShut {NoStop}%
\bibitem [{\citenamefont {Campbell}\ and\ \citenamefont
  {Deffner}(2017)}]{Campbell2017}%
  \BibitemOpen
  \bibfield  {author} {\bibinfo {author} {\bibfnamefont {S.}~\bibnamefont
  {Campbell}}\ and\ \bibinfo {author} {\bibfnamefont {S.}~\bibnamefont
  {Deffner}},\ }\bibfield  {title} {\enquote {\bibinfo {title} {{Trade-Off
  Between Speed and Cost in Shortcuts to Adiabaticity}},}\ }\href {\doibase
  10.1103/PhysRevLett.118.100601} {\bibfield  {journal} {\bibinfo  {journal}
  {Phys. Rev. Lett.}\ }\textbf {\bibinfo {volume} {118}},\ \bibinfo {pages}
  {100601} (\bibinfo {year} {2017})}\BibitemShut {NoStop}%
\bibitem [{\citenamefont {Funo}\ \emph {et~al.}(2017)\citenamefont {Funo},
  \citenamefont {Zhang}, \citenamefont {Chatou}, \citenamefont {Kim},
  \citenamefont {Ueda},\ and\ \citenamefont {del Campo}}]{Funo2017a}%
  \BibitemOpen
  \bibfield  {author} {\bibinfo {author} {\bibfnamefont {K.}~\bibnamefont
  {Funo}}, \bibinfo {author} {\bibfnamefont {J.-N.}\ \bibnamefont {Zhang}},
  \bibinfo {author} {\bibfnamefont {C.}~\bibnamefont {Chatou}}, \bibinfo
  {author} {\bibfnamefont {K.}~\bibnamefont {Kim}}, \bibinfo {author}
  {\bibfnamefont {M.}~\bibnamefont {Ueda}}, \ and\ \bibinfo {author}
  {\bibfnamefont {A.}~\bibnamefont {del Campo}},\ }\bibfield  {title} {\enquote
  {\bibinfo {title} {{Universal Work Fluctuations During Shortcuts to
  Adiabaticity by Counterdiabatic Driving}},}\ }\href {\doibase
  10.1103/PhysRevLett.118.100602} {\bibfield  {journal} {\bibinfo  {journal}
  {Phys. Rev. Lett.}\ }\textbf {\bibinfo {volume} {118}},\ \bibinfo {pages}
  {100602} (\bibinfo {year} {2017})}\BibitemShut {NoStop}%
\bibitem [{\citenamefont {Campaioli}\ \emph
  {et~al.}(2017{\natexlab{a}})\citenamefont {Campaioli}, \citenamefont
  {Pollock}, \citenamefont {Binder}, \citenamefont {C{\'{e}}leri},
  \citenamefont {Goold}, \citenamefont {Vinjanampathy},\ and\ \citenamefont
  {Modi}}]{Campaioli2017}%
  \BibitemOpen
  \bibfield  {author} {\bibinfo {author} {\bibfnamefont {F.}~\bibnamefont
  {Campaioli}}, \bibinfo {author} {\bibfnamefont {F.~A.}\ \bibnamefont
  {Pollock}}, \bibinfo {author} {\bibfnamefont {F.~C.}\ \bibnamefont {Binder}},
  \bibinfo {author} {\bibfnamefont {L.}~\bibnamefont {C{\'{e}}leri}}, \bibinfo
  {author} {\bibfnamefont {J.}~\bibnamefont {Goold}}, \bibinfo {author}
  {\bibfnamefont {S.}~\bibnamefont {Vinjanampathy}}, \ and\ \bibinfo {author}
  {\bibfnamefont {K.}~\bibnamefont {Modi}},\ }\bibfield  {title} {\enquote
  {\bibinfo {title} {{Enhancing the Charging Power of Quantum Batteries}},}\
  }\href {\doibase 10.1103/PhysRevLett.118.150601} {\bibfield  {journal}
  {\bibinfo  {journal} {Phys. Rev. Lett.}\ }\textbf {\bibinfo {volume} {118}},\
  \bibinfo {pages} {150601} (\bibinfo {year} {2017}{\natexlab{a}})}\BibitemShut
  {NoStop}%
\bibitem [{\citenamefont {Campaioli}\ \emph {et~al.}(2018)\citenamefont
  {Campaioli}, \citenamefont {Pollock},\ and\ \citenamefont
  {Vinjanampathy}}]{Campaioli2018a}%
  \BibitemOpen
  \bibfield  {author} {\bibinfo {author} {\bibfnamefont {F.}~\bibnamefont
  {Campaioli}}, \bibinfo {author} {\bibfnamefont {F.~A.}\ \bibnamefont
  {Pollock}}, \ and\ \bibinfo {author} {\bibfnamefont {S.}~\bibnamefont
  {Vinjanampathy}},\ }\enquote {\bibinfo {title} {Quantum batteries},}\ in\
  \href {\doibase 10.1007/978-3-319-99046-0_8} {\emph {\bibinfo {booktitle}
  {Thermodynamics in the Quantum Regime: Fundamental Aspects and New
  Directions}}},\ \bibinfo {editor} {edited by\ \bibinfo {editor}
  {\bibfnamefont {F.}~\bibnamefont {Binder}}, \bibinfo {editor} {\bibfnamefont
  {L.~A.}\ \bibnamefont {Correa}}, \bibinfo {editor} {\bibfnamefont
  {C.}~\bibnamefont {Gogolin}}, \bibinfo {editor} {\bibfnamefont
  {J.}~\bibnamefont {Anders}}, \ and\ \bibinfo {editor} {\bibfnamefont
  {G.}~\bibnamefont {Adesso}}}\ (\bibinfo  {publisher} {Springer International
  Publishing},\ \bibinfo {address} {Cham},\ \bibinfo {year} {2018})\ pp.\
  \bibinfo {pages} {207--225}\BibitemShut {NoStop}%
\bibitem [{\citenamefont {Okuyama}\ and\ \citenamefont
  {Ohzeki}(2018)}]{Okuyama2018}%
  \BibitemOpen
  \bibfield  {author} {\bibinfo {author} {\bibfnamefont {M.}~\bibnamefont
  {Okuyama}}\ and\ \bibinfo {author} {\bibfnamefont {M.}~\bibnamefont
  {Ohzeki}},\ }\bibfield  {title} {\enquote {\bibinfo {title} {{Quantum Speed
  Limit is Not Quantum}},}\ }\href {\doibase 10.1103/PhysRevLett.120.070402}
  {\bibfield  {journal} {\bibinfo  {journal} {Phys. Rev. Lett.}\ }\textbf
  {\bibinfo {volume} {120}},\ \bibinfo {pages} {070402} (\bibinfo {year}
  {2018})}\BibitemShut {NoStop}%
\bibitem [{\citenamefont {Shanahan}\ \emph {et~al.}(2018)\citenamefont
  {Shanahan}, \citenamefont {Chenu}, \citenamefont {Margolus},\ and\
  \citenamefont {del Campo}}]{Shanahan2018}%
  \BibitemOpen
  \bibfield  {author} {\bibinfo {author} {\bibfnamefont {B.}~\bibnamefont
  {Shanahan}}, \bibinfo {author} {\bibfnamefont {A.}~\bibnamefont {Chenu}},
  \bibinfo {author} {\bibfnamefont {N.}~\bibnamefont {Margolus}}, \ and\
  \bibinfo {author} {\bibfnamefont {A.}~\bibnamefont {del Campo}},\ }\bibfield
  {title} {\enquote {\bibinfo {title} {{Quantum Speed Limits across the
  Quantum-to-Classical Transition}},}\ }\href {\doibase
  10.1103/PhysRevLett.120.070401} {\bibfield  {journal} {\bibinfo  {journal}
  {Phys. Rev. Lett.}\ }\textbf {\bibinfo {volume} {120}},\ \bibinfo {pages}
  {070401} (\bibinfo {year} {2018})}\BibitemShut {NoStop}%
\bibitem [{\citenamefont {Kupferman}\ and\ \citenamefont
  {Reznik}(2008)}]{Kupferman2008}%
  \BibitemOpen
  \bibfield  {author} {\bibinfo {author} {\bibfnamefont {J.}~\bibnamefont
  {Kupferman}}\ and\ \bibinfo {author} {\bibfnamefont {B.}~\bibnamefont
  {Reznik}},\ }\bibfield  {title} {\enquote {\bibinfo {title} {{Entanglement
  and the speed of evolution in mixed states}},}\ }\href {\doibase
  10.1103/PhysRevA.78.042305} {\bibfield  {journal} {\bibinfo  {journal} {Phys.
  Rev. A}\ }\textbf {\bibinfo {volume} {78}},\ \bibinfo {pages} {042305}
  (\bibinfo {year} {2008})}\BibitemShut {NoStop}%
\bibitem [{\citenamefont {Uzdin}\ \emph {et~al.}(2012)\citenamefont {Uzdin},
  \citenamefont {G{\"{u}}nther}, \citenamefont {Rahav},\ and\ \citenamefont
  {Moiseyev}}]{Uzdin2012}%
  \BibitemOpen
  \bibfield  {author} {\bibinfo {author} {\bibfnamefont {R.}~\bibnamefont
  {Uzdin}}, \bibinfo {author} {\bibfnamefont {U.}~\bibnamefont
  {G{\"{u}}nther}}, \bibinfo {author} {\bibfnamefont {S.}~\bibnamefont
  {Rahav}}, \ and\ \bibinfo {author} {\bibfnamefont {N.}~\bibnamefont
  {Moiseyev}},\ }\bibfield  {title} {\enquote {\bibinfo {title}
  {{Time-dependent Hamiltonians with 100{\%} evolution speed efficiency}},}\
  }\href {\doibase 10.1088/1751-8113/45/41/415304} {\bibfield  {journal}
  {\bibinfo  {journal} {J. Phys. A Math. Theor.}\ }\textbf {\bibinfo {volume}
  {45}},\ \bibinfo {pages} {415304} (\bibinfo {year} {2012})}\BibitemShut
  {NoStop}%
\bibitem [{\citenamefont {Santos}\ and\ \citenamefont
  {Sarandy}(2015)}]{Santos2015}%
  \BibitemOpen
  \bibfield  {author} {\bibinfo {author} {\bibfnamefont {A.~C.}\ \bibnamefont
  {Santos}}\ and\ \bibinfo {author} {\bibfnamefont {M.~S.}\ \bibnamefont
  {Sarandy}},\ }\bibfield  {title} {\enquote {\bibinfo {title} {{Superadiabatic
  Controlled Evolutions and Universal Quantum Computation}},}\ }\href {\doibase
  10.1038/srep15775} {\bibfield  {journal} {\bibinfo  {journal} {Sci. Rep.}\
  }\textbf {\bibinfo {volume} {5}},\ \bibinfo {pages} {15775} (\bibinfo {year}
  {2015})}\BibitemShut {NoStop}%
\bibitem [{\citenamefont {Santos}\ \emph {et~al.}(2016)\citenamefont {Santos},
  \citenamefont {Silva},\ and\ \citenamefont {Sarandy}}]{Santos2016}%
  \BibitemOpen
  \bibfield  {author} {\bibinfo {author} {\bibfnamefont {A.~C.}\ \bibnamefont
  {Santos}}, \bibinfo {author} {\bibfnamefont {R.~D.}\ \bibnamefont {Silva}}, \
  and\ \bibinfo {author} {\bibfnamefont {M.~S.}\ \bibnamefont {Sarandy}},\
  }\bibfield  {title} {\enquote {\bibinfo {title} {{Shortcut to adiabatic gate
  teleportation}},}\ }\href {\doibase 10.1103/PhysRevA.93.012311} {\bibfield
  {journal} {\bibinfo  {journal} {Phys. Rev. A}\ }\textbf {\bibinfo {volume}
  {93}},\ \bibinfo {pages} {012311} (\bibinfo {year} {2016})}\BibitemShut
  {NoStop}%
\bibitem [{\citenamefont {Goold}\ \emph {et~al.}(2016)\citenamefont {Goold},
  \citenamefont {Huber}, \citenamefont {Riera}, \citenamefont {del Rio},\ and\
  \citenamefont {Skrzypczyk}}]{Goold2016}%
  \BibitemOpen
  \bibfield  {author} {\bibinfo {author} {\bibfnamefont {J.}~\bibnamefont
  {Goold}}, \bibinfo {author} {\bibfnamefont {M.}~\bibnamefont {Huber}},
  \bibinfo {author} {\bibfnamefont {A.}~\bibnamefont {Riera}}, \bibinfo
  {author} {\bibfnamefont {L.}~\bibnamefont {del Rio}}, \ and\ \bibinfo
  {author} {\bibfnamefont {P.}~\bibnamefont {Skrzypczyk}},\ }\bibfield  {title}
  {\enquote {\bibinfo {title} {{The role of quantum information in
  thermodynamics—a topical review}},}\ }\href {\doibase
  10.1088/1751-8113/49/14/143001} {\bibfield  {journal} {\bibinfo  {journal}
  {J. Phys. A Math. Theor.}\ }\textbf {\bibinfo {volume} {49}},\ \bibinfo
  {pages} {143001} (\bibinfo {year} {2016})}\BibitemShut {NoStop}%
\bibitem [{\citenamefont {Uzdin}\ and\ \citenamefont
  {Kosloff}(2016)}]{Uzdin2016}%
  \BibitemOpen
  \bibfield  {author} {\bibinfo {author} {\bibfnamefont {R.}~\bibnamefont
  {Uzdin}}\ and\ \bibinfo {author} {\bibfnamefont {R.}~\bibnamefont
  {Kosloff}},\ }\bibfield  {title} {\enquote {\bibinfo {title} {{Speed limits
  in Liouville space for open quantum systems}},}\ }\href {\doibase
  10.1209/0295-5075/115/40003} {\bibfield  {journal} {\bibinfo  {journal} {EPL
  (Europhysics Lett.}\ }\textbf {\bibinfo {volume} {115}},\ \bibinfo {pages}
  {40003} (\bibinfo {year} {2016})}\BibitemShut {NoStop}%
\bibitem [{\citenamefont {Mondal}\ \emph {et~al.}(2015)\citenamefont {Mondal},
  \citenamefont {Datta},\ and\ \citenamefont {Sazim}}]{Mondal2016}%
  \BibitemOpen
  \bibfield  {author} {\bibinfo {author} {\bibfnamefont {D.}~\bibnamefont
  {Mondal}}, \bibinfo {author} {\bibfnamefont {C.}~\bibnamefont {Datta}}, \
  and\ \bibinfo {author} {\bibfnamefont {S.}~\bibnamefont {Sazim}},\ }\bibfield
   {title} {\enquote {\bibinfo {title} {{Quantum coherence sets the quantum
  speed limit for mixed states}},}\ }\href {\doibase
  10.1016/J.PHYSLETA.2015.12.015} {\bibfield  {journal} {\bibinfo  {journal}
  {Phys. Lett. A}\ }\textbf {\bibinfo {volume} {380}},\ \bibinfo {pages} {689}
  (\bibinfo {year} {2015})}\BibitemShut {NoStop}%
\bibitem [{\citenamefont {Mondal}\ and\ \citenamefont
  {Pati}(2016)}]{Mondal2016b}%
  \BibitemOpen
  \bibfield  {author} {\bibinfo {author} {\bibfnamefont {D.}~\bibnamefont
  {Mondal}}\ and\ \bibinfo {author} {\bibfnamefont {A.~K.}\ \bibnamefont
  {Pati}},\ }\bibfield  {title} {\enquote {\bibinfo {title} {{Quantum speed
  limit for mixed states using an experimentally realizable metric}},}\ }\href
  {\doibase 10.1016/J.PHYSLETA.2016.02.018} {\bibfield  {journal} {\bibinfo
  {journal} {Phys. Lett. A}\ }\textbf {\bibinfo {volume} {380}},\ \bibinfo
  {pages} {1395} (\bibinfo {year} {2016})}\BibitemShut {NoStop}%
\bibitem [{\citenamefont {Mirkin}\ \emph {et~al.}(2016)\citenamefont {Mirkin},
  \citenamefont {Toscano},\ and\ \citenamefont {Wisniacki}}]{Mirkin2016}%
  \BibitemOpen
  \bibfield  {author} {\bibinfo {author} {\bibfnamefont {N.}~\bibnamefont
  {Mirkin}}, \bibinfo {author} {\bibfnamefont {F.}~\bibnamefont {Toscano}}, \
  and\ \bibinfo {author} {\bibfnamefont {D.~A.}\ \bibnamefont {Wisniacki}},\
  }\bibfield  {title} {\enquote {\bibinfo {title} {{Quantum-speed-limit bounds
  in an open quantum evolution}},}\ }\href {\doibase
  10.1103/PhysRevA.94.052125} {\bibfield  {journal} {\bibinfo  {journal} {Phys.
  Rev. A}\ }\textbf {\bibinfo {volume} {94}},\ \bibinfo {pages} {052125}
  (\bibinfo {year} {2016})}\BibitemShut {NoStop}%
\bibitem [{\citenamefont {Pires}\ \emph {et~al.}(2016)\citenamefont {Pires},
  \citenamefont {Cianciaruso}, \citenamefont {C{\'{e}}leri}, \citenamefont
  {Adesso},\ and\ \citenamefont {Soares-Pinto}}]{Pires2016}%
  \BibitemOpen
  \bibfield  {author} {\bibinfo {author} {\bibfnamefont {D.~P.}\ \bibnamefont
  {Pires}}, \bibinfo {author} {\bibfnamefont {M.}~\bibnamefont {Cianciaruso}},
  \bibinfo {author} {\bibfnamefont {L.~C.}\ \bibnamefont {C{\'{e}}leri}},
  \bibinfo {author} {\bibfnamefont {G.}~\bibnamefont {Adesso}}, \ and\ \bibinfo
  {author} {\bibfnamefont {D.~O.}\ \bibnamefont {Soares-Pinto}},\ }\bibfield
  {title} {\enquote {\bibinfo {title} {{Generalized Geometric Quantum Speed
  Limits}},}\ }\href {\doibase 10.1103/PhysRevX.6.021031} {\bibfield  {journal}
  {\bibinfo  {journal} {Phys. Rev. X}\ }\textbf {\bibinfo {volume} {6}},\
  \bibinfo {pages} {021031} (\bibinfo {year} {2016})}\BibitemShut {NoStop}%
\bibitem [{\citenamefont {Marvian}\ \emph {et~al.}(2016)\citenamefont
  {Marvian}, \citenamefont {Spekkens},\ and\ \citenamefont
  {Zanardi}}]{Marvian2016}%
  \BibitemOpen
  \bibfield  {author} {\bibinfo {author} {\bibfnamefont {I.}~\bibnamefont
  {Marvian}}, \bibinfo {author} {\bibfnamefont {R.~W.}\ \bibnamefont
  {Spekkens}}, \ and\ \bibinfo {author} {\bibfnamefont {P.}~\bibnamefont
  {Zanardi}},\ }\bibfield  {title} {\enquote {\bibinfo {title} {{Quantum speed
  limits, coherence, and asymmetry}},}\ }\href {\doibase
  10.1103/PhysRevA.93.052331} {\bibfield  {journal} {\bibinfo  {journal} {Phys.
  Rev. A}\ }\textbf {\bibinfo {volume} {93}},\ \bibinfo {pages} {052331}
  (\bibinfo {year} {2016})}\BibitemShut {NoStop}%
\bibitem [{\citenamefont {Friis}\ \emph {et~al.}(2016)\citenamefont {Friis},
  \citenamefont {Huber},\ and\ \citenamefont {Perarnau-Llobet}}]{Friis2016}%
  \BibitemOpen
  \bibfield  {author} {\bibinfo {author} {\bibfnamefont {N.}~\bibnamefont
  {Friis}}, \bibinfo {author} {\bibfnamefont {M.}~\bibnamefont {Huber}}, \ and\
  \bibinfo {author} {\bibfnamefont {M.}~\bibnamefont {Perarnau-Llobet}},\
  }\bibfield  {title} {\enquote {\bibinfo {title} {{Energetics of correlations
  in interacting systems}},}\ }\href {\doibase 10.1103/PhysRevE.93.042135}
  {\bibfield  {journal} {\bibinfo  {journal} {Phys. Rev. E}\ }\textbf {\bibinfo
  {volume} {93}},\ \bibinfo {pages} {042135} (\bibinfo {year}
  {2016})}\BibitemShut {NoStop}%
\bibitem [{\citenamefont {Epstein}\ and\ \citenamefont
  {Whaley}(2017)}]{Epstein2017}%
  \BibitemOpen
  \bibfield  {author} {\bibinfo {author} {\bibfnamefont {J.~M.}\ \bibnamefont
  {Epstein}}\ and\ \bibinfo {author} {\bibfnamefont {K.~B.}\ \bibnamefont
  {Whaley}},\ }\bibfield  {title} {\enquote {\bibinfo {title} {{Quantum speed
  limits for quantum-information-processing tasks}},}\ }\href {\doibase
  10.1103/PhysRevA.95.042314} {\bibfield  {journal} {\bibinfo  {journal} {Phys.
  Rev. A}\ }\textbf {\bibinfo {volume} {95}},\ \bibinfo {pages} {042314}
  (\bibinfo {year} {2017})}\BibitemShut {NoStop}%
\bibitem [{\citenamefont {Ektesabi}\ \emph {et~al.}(2017)\citenamefont
  {Ektesabi}, \citenamefont {Behzadi},\ and\ \citenamefont
  {Faizi}}]{Ektesabi2017}%
  \BibitemOpen
  \bibfield  {author} {\bibinfo {author} {\bibfnamefont {A.}~\bibnamefont
  {Ektesabi}}, \bibinfo {author} {\bibfnamefont {N.}~\bibnamefont {Behzadi}}, \
  and\ \bibinfo {author} {\bibfnamefont {E.}~\bibnamefont {Faizi}},\ }\bibfield
   {title} {\enquote {\bibinfo {title} {{Improved bound for quantum-speed-limit
  time in open quantum systems by introducing an alternative fidelity}},}\
  }\href {\doibase 10.1103/PhysRevA.95.022115} {\bibfield  {journal} {\bibinfo
  {journal} {Phys. Rev. A}\ }\textbf {\bibinfo {volume} {95}},\ \bibinfo
  {pages} {022115} (\bibinfo {year} {2017})}\BibitemShut {NoStop}%
\bibitem [{\citenamefont {Russell}\ and\ \citenamefont
  {Stepney}(2017)}]{Russell2017}%
  \BibitemOpen
  \bibfield  {author} {\bibinfo {author} {\bibfnamefont {B.}~\bibnamefont
  {Russell}}\ and\ \bibinfo {author} {\bibfnamefont {S.}~\bibnamefont
  {Stepney}},\ }\bibfield  {title} {\enquote {\bibinfo {title} {{The Geometry
  of Speed Limiting Resources in Physical Models of Computation}},}\ }\href
  {\doibase 10.1142/S0129054117500204} {\bibfield  {journal} {\bibinfo
  {journal} {Int. J. Found. Comput. Sci.}\ }\textbf {\bibinfo {volume} {28}},\
  \bibinfo {pages} {321} (\bibinfo {year} {2017})}\BibitemShut {NoStop}%
\bibitem [{\citenamefont {Garc{\'{\i}}a-Pintos}\ and\ \citenamefont {del
  Campo}(2019)}]{Garcia-Pintos2018}%
  \BibitemOpen
  \bibfield  {author} {\bibinfo {author} {\bibfnamefont {L.~P.}\ \bibnamefont
  {Garc{\'{\i}}a-Pintos}}\ and\ \bibinfo {author} {\bibfnamefont
  {A.}~\bibnamefont {del Campo}},\ }\bibfield  {title} {\enquote {\bibinfo
  {title} {Quantum speed limits under continuous quantum measurements},}\
  }\href {\doibase 10.1088/1367-2630/ab099e} {\bibfield  {journal} {\bibinfo
  {journal} {New Journal of Physics}\ }\textbf {\bibinfo {volume} {21}},\
  \bibinfo {pages} {033012} (\bibinfo {year} {2019})}\BibitemShut {NoStop}%
\bibitem [{\citenamefont {Berrada}(2018)}]{Berrada2018}%
  \BibitemOpen
  \bibfield  {author} {\bibinfo {author} {\bibfnamefont {K.}~\bibnamefont
  {Berrada}},\ }\bibfield  {title} {\enquote {\bibinfo {title} {{Quantum
  speedup in structured environments}},}\ }\href {\doibase
  10.1016/J.PHYSE.2017.08.020} {\bibfield  {journal} {\bibinfo  {journal}
  {Phys. E Low-dimensional Syst. Nanostructures}\ }\textbf {\bibinfo {volume}
  {95}},\ \bibinfo {pages} {6} (\bibinfo {year} {2018})}\BibitemShut {NoStop}%
\bibitem [{\citenamefont {Santos}\ and\ \citenamefont
  {Sarandy}(2018)}]{Santos2018}%
  \BibitemOpen
  \bibfield  {author} {\bibinfo {author} {\bibfnamefont {A.~C.}\ \bibnamefont
  {Santos}}\ and\ \bibinfo {author} {\bibfnamefont {M.~S.}\ \bibnamefont
  {Sarandy}},\ }\bibfield  {title} {\enquote {\bibinfo {title} {{Generalized
  shortcuts to adiabaticity and enhanced robustness against decoherence}},}\
  }\href {\doibase 10.1088/1751-8121/aa96f1} {\bibfield  {journal} {\bibinfo
  {journal} {J. Phys. A Math. Theor.}\ }\textbf {\bibinfo {volume} {51}},\
  \bibinfo {pages} {025301} (\bibinfo {year} {2018})}\BibitemShut {NoStop}%
\bibitem [{\citenamefont {Hu}\ \emph {et~al.}(2018)\citenamefont {Hu},
  \citenamefont {Cui}, \citenamefont {Santos}, \citenamefont {Huang},
  \citenamefont {Sarandy}, \citenamefont {Li},\ and\ \citenamefont
  {Guo}}]{Hu2018}%
  \BibitemOpen
  \bibfield  {author} {\bibinfo {author} {\bibfnamefont {C.-K.}\ \bibnamefont
  {Hu}}, \bibinfo {author} {\bibfnamefont {J.-M.}\ \bibnamefont {Cui}},
  \bibinfo {author} {\bibfnamefont {A.~C.}\ \bibnamefont {Santos}}, \bibinfo
  {author} {\bibfnamefont {Y.-F.}\ \bibnamefont {Huang}}, \bibinfo {author}
  {\bibfnamefont {M.~S.}\ \bibnamefont {Sarandy}}, \bibinfo {author}
  {\bibfnamefont {C.-F.}\ \bibnamefont {Li}}, \ and\ \bibinfo {author}
  {\bibfnamefont {G.-C.}\ \bibnamefont {Guo}},\ }\bibfield  {title} {\enquote
  {\bibinfo {title} {{Experimental implementation of generalized transitionless
  quantum driving}},}\ }\href {\doibase 10.1364/OL.43.003136} {\bibfield
  {journal} {\bibinfo  {journal} {Opt. Lett.}\ }\textbf {\bibinfo {volume}
  {43}},\ \bibinfo {pages} {3136} (\bibinfo {year} {2018})}\BibitemShut
  {NoStop}%
\bibitem [{\citenamefont {Volkoff}\ and\ \citenamefont
  {Whaley}(2018)}]{Volkoff2018}%
  \BibitemOpen
  \bibfield  {author} {\bibinfo {author} {\bibfnamefont {T.}~\bibnamefont
  {Volkoff}}\ and\ \bibinfo {author} {\bibfnamefont {K.}~\bibnamefont
  {Whaley}},\ }\bibfield  {title} {\enquote {\bibinfo {title}
  {{Distinguishability times and asymmetry monotone-based quantum speed limits
  in the Bloch ball}},}\ }\href {\doibase 10.22331/q-2018-10-01-96} {\bibfield
  {journal} {\bibinfo  {journal} {Quantum}\ }\textbf {\bibinfo {volume} {2}},\
  \bibinfo {pages} {96} (\bibinfo {year} {2018})}\BibitemShut {NoStop}%
\bibitem [{\citenamefont {Fubini}(1904)}]{Fubini1904}%
  \BibitemOpen
  \bibfield  {author} {\bibinfo {author} {\bibfnamefont {G.}~\bibnamefont
  {Fubini}},\ }\bibfield  {title} {\enquote {\bibinfo {title} {{Sulle metriche
  definite da una forma Hermitiana}},}\ }\href@noop {} {\bibfield  {journal}
  {\bibinfo  {journal} {Atti Istit. Veneto}\ }\textbf {\bibinfo {volume}
  {63}},\ \bibinfo {pages} {502} (\bibinfo {year} {1904})}\BibitemShut
  {NoStop}%
\bibitem [{\citenamefont {Study}(1905)}]{Study1905}%
  \BibitemOpen
  \bibfield  {author} {\bibinfo {author} {\bibfnamefont {E.}~\bibnamefont
  {Study}},\ }\bibfield  {title} {\enquote {\bibinfo {title} {{K{\"{u}}rzeste
  Wege im komplexen Gebiet}},}\ }\href@noop {} {\bibfield  {journal} {\bibinfo
  {journal} {Math. Ann.}\ }\textbf {\bibinfo {volume} {60}},\ \bibinfo {pages}
  {321} (\bibinfo {year} {1905})}\BibitemShut {NoStop}%
\bibitem [{\citenamefont {Bengtsson}\ and\ \citenamefont
  {Zyczkowski}(2008)}]{Bengtsson2008}%
  \BibitemOpen
  \bibfield  {author} {\bibinfo {author} {\bibfnamefont {I.}~\bibnamefont
  {Bengtsson}}\ and\ \bibinfo {author} {\bibfnamefont {K.}~\bibnamefont
  {Zyczkowski}},\ }\href@noop {} {\emph {\bibinfo {title} {{Geometry of quantum
  states : an introduction to quantum entanglement}}}}\ (\bibinfo  {publisher}
  {Cambridge University Press},\ \bibinfo {year} {2008})\ p.\ \bibinfo {pages}
  {419}\BibitemShut {NoStop}%
\bibitem [{\citenamefont {Levitin}\ and\ \citenamefont
  {Toffoli}(2009)}]{Levitin2009}%
  \BibitemOpen
  \bibfield  {author} {\bibinfo {author} {\bibfnamefont {L.~B.}\ \bibnamefont
  {Levitin}}\ and\ \bibinfo {author} {\bibfnamefont {T.}~\bibnamefont
  {Toffoli}},\ }\bibfield  {title} {\enquote {\bibinfo {title} {{Fundamental
  limit on the rate of quantum dynamics: The unified bound is tight}},}\ }\href
  {https://journals.aps.org/prl/abstract/10.1103/PhysRevLett.103.160502}
  {\bibfield  {journal} {\bibinfo  {journal} {Phys. Rev. Lett.}\ }\textbf
  {\bibinfo {volume} {103}},\ \bibinfo {pages} {160502} (\bibinfo {year}
  {2009})}\BibitemShut {NoStop}%
\bibitem [{\citenamefont {Deffner}\ and\ \citenamefont
  {Lutz}(2013{\natexlab{b}})}]{Deffner2013b}%
  \BibitemOpen
  \bibfield  {author} {\bibinfo {author} {\bibfnamefont {S.}~\bibnamefont
  {Deffner}}\ and\ \bibinfo {author} {\bibfnamefont {E.}~\bibnamefont {Lutz}},\
  }\bibfield  {title} {\enquote {\bibinfo {title} {{Quantum Speed Limit for
  Non-Markovian Dynamics}},}\ }\href {\doibase 10.1103/PhysRevLett.111.010402}
  {\bibfield  {journal} {\bibinfo  {journal} {Phys. Rev. Lett}\ }\textbf
  {\bibinfo {volume} {111}},\ \bibinfo {pages} {010402} (\bibinfo {year}
  {2013}{\natexlab{b}})}\BibitemShut {NoStop}%
\bibitem [{\citenamefont {del Campo}\ \emph {et~al.}(2013)\citenamefont {del
  Campo}, \citenamefont {Egusquiza}, \citenamefont {Plenio},\ and\
  \citenamefont {Huelga}}]{DelCampo2013}%
  \BibitemOpen
  \bibfield  {author} {\bibinfo {author} {\bibfnamefont {A.}~\bibnamefont {del
  Campo}}, \bibinfo {author} {\bibfnamefont {I.~L.}\ \bibnamefont {Egusquiza}},
  \bibinfo {author} {\bibfnamefont {M.~B.}\ \bibnamefont {Plenio}}, \ and\
  \bibinfo {author} {\bibfnamefont {S.~F.}\ \bibnamefont {Huelga}},\ }\bibfield
   {title} {\enquote {\bibinfo {title} {{Quantum Speed Limits in Open System
  Dynamics}},}\ }\href {\doibase 10.1103/PhysRevLett.110.050403} {\bibfield
  {journal} {\bibinfo  {journal} {Phys. Rev. Lett.}\ }\textbf {\bibinfo
  {volume} {110}},\ \bibinfo {pages} {050403} (\bibinfo {year}
  {2013})}\BibitemShut {NoStop}%
\bibitem [{\citenamefont {Sun}\ \emph {et~al.}(2015)\citenamefont {Sun},
  \citenamefont {Liu}, \citenamefont {Ma},\ and\ \citenamefont
  {Wang}}]{Sun2015}%
  \BibitemOpen
  \bibfield  {author} {\bibinfo {author} {\bibfnamefont {Z.}~\bibnamefont
  {Sun}}, \bibinfo {author} {\bibfnamefont {J.}~\bibnamefont {Liu}}, \bibinfo
  {author} {\bibfnamefont {J.}~\bibnamefont {Ma}}, \ and\ \bibinfo {author}
  {\bibfnamefont {X.}~\bibnamefont {Wang}},\ }\bibfield  {title} {\enquote
  {\bibinfo {title} {{Quantum speed limits in open systems: Non-Markovian
  dynamics without rotating-wave approximation}},}\ }\href {\doibase
  10.1038/srep08444} {\bibfield  {journal} {\bibinfo  {journal} {Sci. Rep.}\
  }\textbf {\bibinfo {volume} {5}},\ \bibinfo {pages} {8444} (\bibinfo {year}
  {2015})}\BibitemShut {NoStop}%
\bibitem [{\citenamefont {Campaioli}\ \emph
  {et~al.}(2017{\natexlab{b}})\citenamefont {Campaioli}, \citenamefont
  {Pollock}, \citenamefont {Binder},\ and\ \citenamefont
  {Modi}}]{Campaioli2018}%
  \BibitemOpen
  \bibfield  {author} {\bibinfo {author} {\bibfnamefont {F.}~\bibnamefont
  {Campaioli}}, \bibinfo {author} {\bibfnamefont {F.~A.}\ \bibnamefont
  {Pollock}}, \bibinfo {author} {\bibfnamefont {F.~C.}\ \bibnamefont {Binder}},
  \ and\ \bibinfo {author} {\bibfnamefont {K.}~\bibnamefont {Modi}},\
  }\bibfield  {title} {\enquote {\bibinfo {title} {{Tightening Quantum Speed
  Limits for Almost All States}},}\ }\href {\doibase
  10.1103/PhysRevLett.120.060409} {\bibfield  {journal} {\bibinfo  {journal}
  {Phys. Rev. Lett.}\ }\textbf {\bibinfo {volume} {120}},\ \bibinfo {pages}
  {060409} (\bibinfo {year} {2017}{\natexlab{b}})}\BibitemShut {NoStop}%
\bibitem [{\citenamefont {Keyl}\ and\ \citenamefont {Werner}(2001)}]{Keyl2001}%
  \BibitemOpen
  \bibfield  {author} {\bibinfo {author} {\bibfnamefont {M.}~\bibnamefont
  {Keyl}}\ and\ \bibinfo {author} {\bibfnamefont {R.~F.}\ \bibnamefont
  {Werner}},\ }\bibfield  {title} {\enquote {\bibinfo {title} {{Estimating the
  spectrum of a density operator}},}\ }\href {\doibase
  10.1103/PhysRevA.64.052311} {\bibfield  {journal} {\bibinfo  {journal} {Phys.
  Rev. A}\ }\textbf {\bibinfo {volume} {64}},\ \bibinfo {pages} {052311}
  (\bibinfo {year} {2001})}\BibitemShut {NoStop}%
\bibitem [{\citenamefont {Ekert}\ \emph {et~al.}(2002)\citenamefont {Ekert},
  \citenamefont {Alves}, \citenamefont {Oi}, \citenamefont {Horodecki},
  \citenamefont {Horodecki},\ and\ \citenamefont {Kwek}}]{Ekert2002}%
  \BibitemOpen
  \bibfield  {author} {\bibinfo {author} {\bibfnamefont {A.~K.}\ \bibnamefont
  {Ekert}}, \bibinfo {author} {\bibfnamefont {C.~M.}\ \bibnamefont {Alves}},
  \bibinfo {author} {\bibfnamefont {D.~K.~L.}\ \bibnamefont {Oi}}, \bibinfo
  {author} {\bibfnamefont {M.}~\bibnamefont {Horodecki}}, \bibinfo {author}
  {\bibfnamefont {P.}~\bibnamefont {Horodecki}}, \ and\ \bibinfo {author}
  {\bibfnamefont {L.~C.}\ \bibnamefont {Kwek}},\ }\bibfield  {title} {\enquote
  {\bibinfo {title} {{Direct Estimations of Linear and Nonlinear Functionals of
  a Quantum State}},}\ }\href {\doibase 10.1103/PhysRevLett.88.217901}
  {\bibfield  {journal} {\bibinfo  {journal} {Phys. Rev. Lett.}\ }\textbf
  {\bibinfo {volume} {88}},\ \bibinfo {pages} {217901} (\bibinfo {year}
  {2002})}\BibitemShut {NoStop}%
\bibitem [{\citenamefont {Russell}\ and\ \citenamefont
  {Stepney}(2014)}]{Russell2014a}%
  \BibitemOpen
  \bibfield  {author} {\bibinfo {author} {\bibfnamefont {B.}~\bibnamefont
  {Russell}}\ and\ \bibinfo {author} {\bibfnamefont {S.}~\bibnamefont
  {Stepney}},\ }\bibfield  {title} {\enquote {\bibinfo {title} {{Applications
  of Finsler Geometry to Speed Limits to Quantum Information Processing}},}\
  }\href {\doibase 10.1142/s0129054114400073} {\bibfield  {journal} {\bibinfo
  {journal} {Int. J. Found. Comput. Sci.}\ }\textbf {\bibinfo {volume} {25}},\
  \bibinfo {pages} {489} (\bibinfo {year} {2014})}\BibitemShut {NoStop}%
\bibitem [{\citenamefont {Wootters}(1981)}]{Wootters1981}%
  \BibitemOpen
  \bibfield  {author} {\bibinfo {author} {\bibfnamefont {W.~K.}\ \bibnamefont
  {Wootters}},\ }\bibfield  {title} {\enquote {\bibinfo {title} {{Statistical
  distance and Hilbert space}},}\ }\href {\doibase 10.1103/PhysRevD.23.357}
  {\bibfield  {journal} {\bibinfo  {journal} {Phys. Rev. D}\ }\textbf {\bibinfo
  {volume} {23}},\ \bibinfo {pages} {357} (\bibinfo {year} {1981})}\BibitemShut
  {NoStop}%
\bibitem [{\citenamefont {Byrd}\ and\ \citenamefont
  {Khaneja}(2003)}]{Byrd2003}%
  \BibitemOpen
  \bibfield  {author} {\bibinfo {author} {\bibfnamefont {M.~S.}\ \bibnamefont
  {Byrd}}\ and\ \bibinfo {author} {\bibfnamefont {N.}~\bibnamefont {Khaneja}},\
  }\bibfield  {title} {\enquote {\bibinfo {title} {{Characterization of the
  Positivity of the Density Matrix in Terms of the Coherence Vector
  Representation}},}\ }\href {\doibase 10.1103/PhysRevA.68.062322} {\bibfield
  {journal} {\bibinfo  {journal} {Phys. Rev. A}\ }\textbf {\bibinfo {volume}
  {68}},\ \bibinfo {pages} {062322} (\bibinfo {year} {2003})}\BibitemShut
  {NoStop}%
\bibitem [{\citenamefont {Taddei}\ \emph {et~al.}(2013)\citenamefont {Taddei},
  \citenamefont {Escher}, \citenamefont {Davidovich},\ and\ \citenamefont {{De
  Matos Filho}}}]{Taddei2013}%
  \BibitemOpen
  \bibfield  {author} {\bibinfo {author} {\bibfnamefont {M.~M.}\ \bibnamefont
  {Taddei}}, \bibinfo {author} {\bibfnamefont {B.~M.}\ \bibnamefont {Escher}},
  \bibinfo {author} {\bibfnamefont {L.}~\bibnamefont {Davidovich}}, \ and\
  \bibinfo {author} {\bibfnamefont {R.~L.}\ \bibnamefont {{De Matos Filho}}},\
  }\bibfield  {title} {\enquote {\bibinfo {title} {{Quantum speed limit for
  physical processes}},}\ }\href {\doibase 10.1103/PhysRevLett.110.050402}
  {\bibfield  {journal} {\bibinfo  {journal} {Phys. Rev. Lett.}\ }\textbf
  {\bibinfo {volume} {110}},\ \bibinfo {pages} {050402} (\bibinfo {year}
  {2013})}\BibitemShut {NoStop}%
\bibitem [{\citenamefont {P{\'{e}}rez-Garc{\'{i}}a}\ \emph
  {et~al.}(2006)\citenamefont {P{\'{e}}rez-Garc{\'{i}}a}, \citenamefont {Wolf},
  \citenamefont {Petz},\ and\ \citenamefont {Ruskai}}]{Perez-Garcia2006}%
  \BibitemOpen
  \bibfield  {author} {\bibinfo {author} {\bibfnamefont {D.}~\bibnamefont
  {P{\'{e}}rez-Garc{\'{i}}a}}, \bibinfo {author} {\bibfnamefont {M.~M.}\
  \bibnamefont {Wolf}}, \bibinfo {author} {\bibfnamefont {D.}~\bibnamefont
  {Petz}}, \ and\ \bibinfo {author} {\bibfnamefont {M.~B.}\ \bibnamefont
  {Ruskai}},\ }\bibfield  {title} {\enquote {\bibinfo {title} {{Contractivity
  of positive and trace-preserving maps under Lp norms}},}\ }\href {\doibase
  10.1063/1.2218675} {\bibfield  {journal} {\bibinfo  {journal} {J. Math.
  Phys.}\ }\textbf {\bibinfo {volume} {47}},\ \bibinfo {pages} {083506}
  (\bibinfo {year} {2006})}\BibitemShut {NoStop}%
\bibitem [{\citenamefont {Piani}(2012)}]{Piani2012}%
  \BibitemOpen
  \bibfield  {author} {\bibinfo {author} {\bibfnamefont {M.}~\bibnamefont
  {Piani}},\ }\bibfield  {title} {\enquote {\bibinfo {title} {{Problem with
  geometric discord}},}\ }\href {\doibase 10.1103/PhysRevA.86.034101}
  {\bibfield  {journal} {\bibinfo  {journal} {Phys. Rev. A}\ }\textbf {\bibinfo
  {volume} {86}},\ \bibinfo {pages} {034101} (\bibinfo {year}
  {2012})}\BibitemShut {NoStop}%
\bibitem [{\citenamefont {Il'ichev}\ \emph {et~al.}(2003)\citenamefont
  {Il'ichev}, \citenamefont {Oukhanski}, \citenamefont {Izmalkov},
  \citenamefont {Wagner}, \citenamefont {Grajcar}, \citenamefont {Meyer},
  \citenamefont {Smirnov}, \citenamefont {{Maassen van den Brink}},
  \citenamefont {Amin},\ and\ \citenamefont {Zagoskin}}]{Ilichev2003}%
  \BibitemOpen
  \bibfield  {author} {\bibinfo {author} {\bibfnamefont {E.}~\bibnamefont
  {Il'ichev}}, \bibinfo {author} {\bibfnamefont {N.}~\bibnamefont {Oukhanski}},
  \bibinfo {author} {\bibfnamefont {A.}~\bibnamefont {Izmalkov}}, \bibinfo
  {author} {\bibfnamefont {T.}~\bibnamefont {Wagner}}, \bibinfo {author}
  {\bibfnamefont {M.}~\bibnamefont {Grajcar}}, \bibinfo {author} {\bibfnamefont
  {H.-G.}\ \bibnamefont {Meyer}}, \bibinfo {author} {\bibfnamefont {A.~Y.}\
  \bibnamefont {Smirnov}}, \bibinfo {author} {\bibfnamefont {A.}~\bibnamefont
  {{Maassen van den Brink}}}, \bibinfo {author} {\bibfnamefont {M.~H.~S.}\
  \bibnamefont {Amin}}, \ and\ \bibinfo {author} {\bibfnamefont {A.~M.}\
  \bibnamefont {Zagoskin}},\ }\bibfield  {title} {\enquote {\bibinfo {title}
  {{Continuous Monitoring of Rabi Oscillations in a Josephson Flux Qubit}},}\
  }\href {\doibase 10.1103/PhysRevLett.91.097906} {\bibfield  {journal}
  {\bibinfo  {journal} {Phys. Rev. Lett.}\ }\textbf {\bibinfo {volume} {91}},\
  \bibinfo {pages} {097906} (\bibinfo {year} {2003})}\BibitemShut {NoStop}%
\bibitem [{\citenamefont {Zueco}\ \emph {et~al.}(2009)\citenamefont {Zueco},
  \citenamefont {Reuther}, \citenamefont {Kohler},\ and\ \citenamefont
  {H{\"{a}}nggi}}]{Zueco2009}%
  \BibitemOpen
  \bibfield  {author} {\bibinfo {author} {\bibfnamefont {D.}~\bibnamefont
  {Zueco}}, \bibinfo {author} {\bibfnamefont {G.~M.}\ \bibnamefont {Reuther}},
  \bibinfo {author} {\bibfnamefont {S.}~\bibnamefont {Kohler}}, \ and\ \bibinfo
  {author} {\bibfnamefont {P.}~\bibnamefont {H{\"{a}}nggi}},\ }\bibfield
  {title} {\enquote {\bibinfo {title} {{Qubit-oscillator dynamics in the
  dispersive regime: Analytical theory beyond the rotating-wave
  approximation}},}\ }\href {\doibase 10.1103/PhysRevA.80.033846} {\bibfield
  {journal} {\bibinfo  {journal} {Phys. Rev. A}\ }\textbf {\bibinfo {volume}
  {80}},\ \bibinfo {pages} {033846} (\bibinfo {year} {2009})}\BibitemShut
  {NoStop}%
\bibitem [{\citenamefont {Rodr{\'{i}}guez-Rosario}\ \emph
  {et~al.}(2011)\citenamefont {Rodr{\'{i}}guez-Rosario}, \citenamefont
  {Kimura}, \citenamefont {Imai},\ and\ \citenamefont
  {Aspuru-Guzik}}]{Rodriguez-Rosario2011}%
  \BibitemOpen
  \bibfield  {author} {\bibinfo {author} {\bibfnamefont {C.~A.}\ \bibnamefont
  {Rodr{\'{i}}guez-Rosario}}, \bibinfo {author} {\bibfnamefont
  {G.}~\bibnamefont {Kimura}}, \bibinfo {author} {\bibfnamefont
  {H.}~\bibnamefont {Imai}}, \ and\ \bibinfo {author} {\bibfnamefont
  {A.}~\bibnamefont {Aspuru-Guzik}},\ }\bibfield  {title} {\enquote {\bibinfo
  {title} {{Sufficient and Necessary Condition for Zero Quantum Entropy Rates
  under any Coupling to the Environment}},}\ }\href {\doibase
  10.1103/PhysRevLett.106.050403} {\bibfield  {journal} {\bibinfo  {journal}
  {Phys. Rev. Lett.}\ }\textbf {\bibinfo {volume} {106}},\ \bibinfo {pages}
  {050403} (\bibinfo {year} {2011})}\BibitemShut {NoStop}%
\bibitem [{\citenamefont {Breuer}\ and\ \citenamefont
  {Petruccione}(2002)}]{Breuer2002}%
  \BibitemOpen
  \bibfield  {author} {\bibinfo {author} {\bibfnamefont {H.-P.}\ \bibnamefont
  {Breuer}}\ and\ \bibinfo {author} {\bibfnamefont {F.~F.}\ \bibnamefont
  {Petruccione}},\ }\href@noop {} {\emph {\bibinfo {title} {{The theory of open
  quantum systems}}}}\ (\bibinfo  {publisher} {Oxford University Press},\
  \bibinfo {year} {2002})\ p.\ \bibinfo {pages} {625}\BibitemShut {NoStop}%
\bibitem [{\citenamefont {Pollock}\ and\ \citenamefont
  {Modi}(2017)}]{Pollock2017}%
  \BibitemOpen
  \bibfield  {author} {\bibinfo {author} {\bibfnamefont {F.~A.}\ \bibnamefont
  {Pollock}}\ and\ \bibinfo {author} {\bibfnamefont {K.}~\bibnamefont {Modi}},\
  }\bibfield  {title} {\enquote {\bibinfo {title} {{Tomographically
  reconstructed master equations for any open quantum dynamics}},}\ }\href
  {\doibase 10.22331/q-2018-07-11-76} {\bibfield  {journal} {\bibinfo
  {journal} {Quantum}\ }\textbf {\bibinfo {volume} {2}},\ \bibinfo {pages} {76}
  (\bibinfo {year} {2017})}\BibitemShut {NoStop}%
\bibitem [{\citenamefont {Uhlmann}(1992)}]{Uhlmann1992b}%
  \BibitemOpen
  \bibfield  {author} {\bibinfo {author} {\bibfnamefont {A.}~\bibnamefont
  {Uhlmann}},\ }\bibfield  {title} {\enquote {\bibinfo {title} {{An energy
  dispersion estimate}},}\ }\href {\doibase 10.1016/0375-9601(92)90555-Z}
  {\bibfield  {journal} {\bibinfo  {journal} {Phys. Lett. A}\ }\textbf
  {\bibinfo {volume} {161}},\ \bibinfo {pages} {329} (\bibinfo {year}
  {1992})}\BibitemShut {NoStop}%
\bibitem [{\citenamefont {Luo}\ and\ \citenamefont {Zhang}(2004)}]{Luo2004}%
  \BibitemOpen
  \bibfield  {author} {\bibinfo {author} {\bibfnamefont {S.}~\bibnamefont
  {Luo}}\ and\ \bibinfo {author} {\bibfnamefont {Q.}~\bibnamefont {Zhang}},\
  }\bibfield  {title} {\enquote {\bibinfo {title} {{Informational distance on
  quantum-state space}},}\ }\href {\doibase 10.1103/PhysRevA.69.032106}
  {\bibfield  {journal} {\bibinfo  {journal} {Phys. Rev. A}\ }\textbf {\bibinfo
  {volume} {69}},\ \bibinfo {pages} {032106} (\bibinfo {year}
  {2004})}\BibitemShut {NoStop}%
\bibitem [{\citenamefont {Facchi}\ \emph {et~al.}(2010)\citenamefont {Facchi},
  \citenamefont {Kulkarni}, \citenamefont {Man'ko}, \citenamefont {Marmo},
  \citenamefont {Sudarshan},\ and\ \citenamefont {Ventriglia}}]{Facchi2010}%
  \BibitemOpen
  \bibfield  {author} {\bibinfo {author} {\bibfnamefont {P.}~\bibnamefont
  {Facchi}}, \bibinfo {author} {\bibfnamefont {R.}~\bibnamefont {Kulkarni}},
  \bibinfo {author} {\bibfnamefont {V.}~\bibnamefont {Man'ko}}, \bibinfo
  {author} {\bibfnamefont {G.}~\bibnamefont {Marmo}}, \bibinfo {author}
  {\bibfnamefont {E.}~\bibnamefont {Sudarshan}}, \ and\ \bibinfo {author}
  {\bibfnamefont {F.}~\bibnamefont {Ventriglia}},\ }\bibfield  {title}
  {\enquote {\bibinfo {title} {{Classical and quantum Fisher information in the
  geometrical formulation of quantum mechanics}},}\ }\href {\doibase
  10.1016/J.PHYSLETA.2010.10.005} {\bibfield  {journal} {\bibinfo  {journal}
  {Phys. Lett. A}\ }\textbf {\bibinfo {volume} {374}},\ \bibinfo {pages} {4801}
  (\bibinfo {year} {2010})}\BibitemShut {NoStop}%
\bibitem [{\citenamefont {Miszczak}\ \emph {et~al.}(2009)\citenamefont
  {Miszczak}, \citenamefont {Pucha{\l}a}, \citenamefont {Horodecki},
  \citenamefont {Uhlmann},\ and\ \citenamefont
  {{\.{Z}}yczkowski}}]{Miszczak2008}%
  \BibitemOpen
  \bibfield  {author} {\bibinfo {author} {\bibfnamefont {J.~A.}\ \bibnamefont
  {Miszczak}}, \bibinfo {author} {\bibfnamefont {Z.}~\bibnamefont
  {Pucha{\l}a}}, \bibinfo {author} {\bibfnamefont {P.}~\bibnamefont
  {Horodecki}}, \bibinfo {author} {\bibfnamefont {A.}~\bibnamefont {Uhlmann}},
  \ and\ \bibinfo {author} {\bibfnamefont {K.}~\bibnamefont
  {{\.{Z}}yczkowski}},\ }\bibfield  {title} {\enquote {\bibinfo {title} {{Sub--
  and super--fidelity as bounds for quantum fidelity}},}\ }\href
  {http://arxiv.org/abs/0805.2037} {\bibfield  {journal} {\bibinfo  {journal}
  {Quantum Inf. Comput.}\ }\textbf {\bibinfo {volume} {9}} (\bibinfo {year}
  {2009})},\ \Eprint {http://arxiv.org/abs/0805.2037} {arXiv:0805.2037}
  \BibitemShut {NoStop}%
\bibitem [{\citenamefont {Abernethy}\ \emph {et~al.}(2009)\citenamefont
  {Abernethy}, \citenamefont {Bach},\ and\ \citenamefont
  {Evgeniou}}]{Abernethy2009}%
  \BibitemOpen
  \bibfield  {author} {\bibinfo {author} {\bibfnamefont {J.}~\bibnamefont
  {Abernethy}}, \bibinfo {author} {\bibfnamefont {F.}~\bibnamefont {Bach}}, \
  and\ \bibinfo {author} {\bibfnamefont {T.}~\bibnamefont {Evgeniou}},\
  }\bibfield  {title} {\enquote {\bibinfo {title} {{A new approach to
  collaborative filtering: Operator estimation with spectral
  regularization}},}\ }\href {http://www.jmlr.org/papers/v10/abernethy09a.html}
  {\bibfield  {journal} {\bibinfo  {journal} {J. Mach. Learn. Res.}\ }\textbf
  {\bibinfo {volume} {10}},\ \bibinfo {pages} {803} (\bibinfo {year}
  {2009})}\BibitemShut {NoStop}%
\bibitem [{\citenamefont {Wang}\ \emph {et~al.}(2015)\citenamefont {Wang},
  \citenamefont {Allegra}, \citenamefont {Jacobs}, \citenamefont {Lloyd},
  \citenamefont {Lupo},\ and\ \citenamefont {Mohseni}}]{Wang2015}%
  \BibitemOpen
  \bibfield  {author} {\bibinfo {author} {\bibfnamefont {X.}~\bibnamefont
  {Wang}}, \bibinfo {author} {\bibfnamefont {M.}~\bibnamefont {Allegra}},
  \bibinfo {author} {\bibfnamefont {K.}~\bibnamefont {Jacobs}}, \bibinfo
  {author} {\bibfnamefont {S.}~\bibnamefont {Lloyd}}, \bibinfo {author}
  {\bibfnamefont {C.}~\bibnamefont {Lupo}}, \ and\ \bibinfo {author}
  {\bibfnamefont {M.}~\bibnamefont {Mohseni}},\ }\bibfield  {title} {\enquote
  {\bibinfo {title} {{Quantum Brachistochrone Curves as Geodesics: Obtaining
  Accurate Minimum-Time Protocols for the Control of Quantum Systems}},}\
  }\href {\doibase 10.1103/PhysRevLett.114.170501} {\bibfield  {journal}
  {\bibinfo  {journal} {Phys. Rev. Lett.}\ }\textbf {\bibinfo {volume} {114}},\
  \bibinfo {pages} {170501} (\bibinfo {year} {2015})}\BibitemShut {NoStop}%
\bibitem [{\citenamefont {Geng}\ \emph {et~al.}(2016)\citenamefont {Geng},
  \citenamefont {Wu}, \citenamefont {Wang}, \citenamefont {Xu}, \citenamefont
  {Shi}, \citenamefont {Xie}, \citenamefont {Rong},\ and\ \citenamefont
  {Du}}]{Geng2016}%
  \BibitemOpen
  \bibfield  {author} {\bibinfo {author} {\bibfnamefont {J.}~\bibnamefont
  {Geng}}, \bibinfo {author} {\bibfnamefont {Y.}~\bibnamefont {Wu}}, \bibinfo
  {author} {\bibfnamefont {X.}~\bibnamefont {Wang}}, \bibinfo {author}
  {\bibfnamefont {K.}~\bibnamefont {Xu}}, \bibinfo {author} {\bibfnamefont
  {F.}~\bibnamefont {Shi}}, \bibinfo {author} {\bibfnamefont {Y.}~\bibnamefont
  {Xie}}, \bibinfo {author} {\bibfnamefont {X.}~\bibnamefont {Rong}}, \ and\
  \bibinfo {author} {\bibfnamefont {J.}~\bibnamefont {Du}},\ }\bibfield
  {title} {\enquote {\bibinfo {title} {{Experimental Time-Optimal Universal
  Control of Spin Qubits in Solids}},}\ }\href {\doibase
  10.1103/PhysRevLett.117.170501} {\bibfield  {journal} {\bibinfo  {journal}
  {Phys. Rev. Lett.}\ }\textbf {\bibinfo {volume} {117}},\ \bibinfo {pages}
  {170501} (\bibinfo {year} {2016})}\BibitemShut {NoStop}%
\bibitem [{\citenamefont {Arenz}\ \emph {et~al.}(2014)\citenamefont {Arenz},
  \citenamefont {Gualdi},\ and\ \citenamefont {Burgarth}}]{Arenz_2014}%
  \BibitemOpen
  \bibfield  {author} {\bibinfo {author} {\bibfnamefont {C.}~\bibnamefont
  {Arenz}}, \bibinfo {author} {\bibfnamefont {G.}~\bibnamefont {Gualdi}}, \
  and\ \bibinfo {author} {\bibfnamefont {D.}~\bibnamefont {Burgarth}},\
  }\bibfield  {title} {\enquote {\bibinfo {title} {Control of open quantum
  systems: case study of the central spin model},}\ }\href {\doibase
  10.1088/1367-2630/16/6/065023} {\bibfield  {journal} {\bibinfo  {journal}
  {New Journal of Physics}\ }\textbf {\bibinfo {volume} {16}},\ \bibinfo
  {pages} {065023} (\bibinfo {year} {2014})}\BibitemShut {NoStop}%
\bibitem [{\citenamefont {Lee}\ \emph {et~al.}(2018)\citenamefont {Lee},
  \citenamefont {Arenz}, \citenamefont {Rabitz},\ and\ \citenamefont
  {Russell}}]{Lee2018}%
  \BibitemOpen
  \bibfield  {author} {\bibinfo {author} {\bibfnamefont {J.}~\bibnamefont
  {Lee}}, \bibinfo {author} {\bibfnamefont {C.}~\bibnamefont {Arenz}}, \bibinfo
  {author} {\bibfnamefont {H.}~\bibnamefont {Rabitz}}, \ and\ \bibinfo {author}
  {\bibfnamefont {B.}~\bibnamefont {Russell}},\ }\bibfield  {title} {\enquote
  {\bibinfo {title} {Dependence of the quantum speed limit on system size and
  control complexity},}\ }\href {\doibase 10.1088/1367-2630/aac6f3} {\bibfield
  {journal} {\bibinfo  {journal} {New Journal of Physics}\ }\textbf {\bibinfo
  {volume} {20}},\ \bibinfo {pages} {063002} (\bibinfo {year}
  {2018})}\BibitemShut {NoStop}%
\bibitem [{\citenamefont {Arenz}\ \emph {et~al.}(2017)\citenamefont {Arenz},
  \citenamefont {Russell}, \citenamefont {Burgarth},\ and\ \citenamefont
  {Rabitz}}]{Arenz_2017}%
  \BibitemOpen
  \bibfield  {author} {\bibinfo {author} {\bibfnamefont {C.}~\bibnamefont
  {Arenz}}, \bibinfo {author} {\bibfnamefont {B.}~\bibnamefont {Russell}},
  \bibinfo {author} {\bibfnamefont {D.}~\bibnamefont {Burgarth}}, \ and\
  \bibinfo {author} {\bibfnamefont {H.}~\bibnamefont {Rabitz}},\ }\bibfield
  {title} {\enquote {\bibinfo {title} {The roles of drift and control field
  constraints upon quantum control speed limits},}\ }\href {\doibase
  10.1088/1367-2630/aa8242} {\bibfield  {journal} {\bibinfo  {journal} {New
  Journal of Physics}\ }\textbf {\bibinfo {volume} {19}},\ \bibinfo {pages}
  {103015} (\bibinfo {year} {2017})}\BibitemShut {NoStop}%
\end{thebibliography}%


%

\newpage
\clearpage
\newpage
\appendix
\section{Derivation of bound~\ref{eq:speed_limit_arbitrary} from distance~\ref{eq:distance}}
\label{a:derivation}
\noindent
Given two states $\rho$, $\sigma\in\mathcal{S}(\mathcal{H}_S)$ of the system, with associated generalized Bloch vectors $\bm{r}$, $\bm{s}$, respectively, the function $D(\rho,\sigma)=\lVert \bm{r}-\bm{s}\rVert_2$ expressed in Eq.~\eqref{eq:distance} is clearly a distance, as it is the Euclidean norm of the displacement vector $\bm{r}-\bm{s}$~\cite{Bengtsson2008}. $D$ can be expressed as a function of the dimension $d$ of the system and of the density matrices $\rho$ and $\sigma$, remembering that
\begin{gather}
    \label{eq:euclidean_hilbertschmidt}
    \begin{split}
    \tr[(\rho-\sigma)^2] & = \tr[(\frac{c}{d} \sum_a (r_a-s_a)\Lambda_a)^2] \\
    & = \frac{d(d-1)}{2d^2}\sum_{a,b}(r_a -s_a)(r_b-s_b)\tr[\Lambda_a\Lambda_b], \\
    & =  \frac{d-1}{2d}\sum_{a,b}(r_a -s_a)(r_b-s_b) 2\delta_{ab}, \\
    & =  \frac{d-1}{d}\sum_{a}(r_a -s_a)^2, \\
    & =  \frac{d-1}{d}\lVert \bm{r} - \bm{s} \rVert_2^2; \\
    \end{split}
\end{gather}
thus, recalling that $\lVert \rho \rVert  = \sqrt{\tr[\rho^\dagger\rho]} = \sqrt{\tr[\rho^2]}$,
\begin{gather}
    D(\rho,\sigma) = \sqrt{\frac{d}{d-1}}\lVert \rho -\sigma \rVert .
\end{gather}
The proof for the QSL bound of Eq.~\eqref{eq:speed_limit_arbitrary} is carried out as follows: Consider a parametric curve $\gamma(s):[0,S]\in\mathbb{R}\to\mathbb{R}^N$ for some $N\geq1$, that connects two different points $A=\gamma(0)$ and $B=\gamma(S)$. Let $\lVert \cdot \rVert_\eta$ be some norm, specified by $\eta$, on $\mathbb{R}^N$, which induces the distance $D(A,B)=\lVert A-B\rVert_\eta$. The length of the path $\gamma$ is given by $L[\gamma_A^B]=\int_0^S ds \lVert \dot{\gamma}(s)\rVert_\eta$, where $\dot{\gamma}(s) = d\gamma(s)/ds$. Since $D(A,B)$ is the geodesic distance between $A$ and $B$, any other path between the two points can be either longer or equal, with respect to the chosen distance (associated by the chosen norm), thus $D(A,B)\leq L[\gamma_A^B]$. In particular, we choose $D(\rho,\sigma)=\lVert \bm{r}-\bm{s}\rVert_2$ and $\bm{r}(t)$ as the parametric curve generated by some arbitrary process, with $\bm{r}(0)=\bm{r}$, and $\bm{r}(\tau)=\bm{s}$. Accordingly, the length of the curve is given by $L[\gamma_{\rho}^{\sigma}] = \int_0^\tau dt \lVert \dot{\bm{r}}(t)\rVert_2$. Recalling that $\lVert \dot{\bm{r}}(t)\rVert_2 = \sqrt{d/(d-1)}\lVert \dot{\rho}_t\rVert $, we obtain the bound. \hfill $\square$

\noindent

\section{Comparison of significant QSL bounds}
\label{a:tightness}
\noindent
As mentioned in the letter, we have considered some significant bounds~\cite{Sun2015,DelCampo2013,Deffner2013b,Mondal2016,Pires2016} to test the performance of our bound $T_D$.

First, we notice Pires's bound is in fact an infinite family of bounds, which depend on the choice of the distance/metric chose to fit the specific type of evolution. Optimal choices of the distance are well known for some notable cases, such as that unitary evolution of pure states, as mentioned above. However, for the general case of arbitrary processes a \emph{preferred} distance has not been specified by authors in~\cite{Pires2016}. For this reason we cannot perform a direct comparison between bound $T_D$ and Pires's, which will be disregarded henceforth.
We then analytically compare our bound to Sun's, Del Campo's, and Deffner's bounds. The last three bounds are given by
\begin{align}
    \label{eq:sun}
    &T_{\textrm{Sun}} = \frac{\bigg|1-\frac{\tr[\rho\sigma]}{\sqrt{\tr[\rho^2]\tr[\sigma^2]}}\bigg|}{2 \overline{\left({\lVert \dot{\rho}_t\rVert }/{\lVert \rho_t \rVert }\right)}},\\
    \label{eq:delcampo} 
    &T_{\textrm{Del Campo}} = \frac{\bigg|1-\frac{\tr[\rho\sigma]}{\tr[\rho^2]}\bigg|\lVert \rho \rVert ^2}{\overline{\lVert \dot{\rho}_t\rVert} },\\
    \label{eq:deffner}
    &T_{\textrm{Deffner}} = \frac{\sin^2\bigg[\arccos \bigg( F(\rho,\sigma)\bigg) \bigg]}{\overline{\lVert \dot{\rho}_t\rVert} },
\end{align}
where $F(\rho,\sigma) = \tr[\sqrt{\sqrt{\rho}\sigma\sqrt{\rho}}]$ is the quantum fidelity between $\rho$ and $\sigma$.
The orbit-dependent term of all of these bounds only depends on the strength of the generator $\lVert \dot{\rho}_t \rVert $, or can be bounded by some quantity that only depends on this term. This observation allows us to evaluate the relative tightness of these bounds and of $T_D$ just by comparing their orbit-independent terms. Let us assume that $\tr[\rho^2]\geq\tr[\sigma^2]$, without loss of generality, and introduce the enhanced bounds
\begin{align}
    \label{eq:over_sun}
    &T_{\textrm{Sun}}^{\;\star} = \frac{\bigg|1-\frac{\tr[\rho\sigma]}{\sqrt{\tr[\rho^2]\tr[\sigma^2]}}\bigg|\lVert \rho \rVert }{2 \overline{\lVert \dot{\rho}_t\rVert} }, \\
    \label{eq:over_deffner}
    &T_{\textrm{Deffner}}^{\;\star} = \frac{\sin^2\bigg[\arccos \bigg(E(\rho,\sigma)\bigg) \bigg]}{\overline{\lVert \dot{\rho}_t\rVert} },
\end{align}
where $E$ is the sub-fidelity
\begin{gather}
    \label{eq:sub_fidelity}
    E(\rho,\sigma)=\sqrt{\tr[\rho\sigma]+\sqrt{2(\tr[\rho\sigma]^2 -\tr[\rho\sigma\rho\sigma]) }},
\end{gather}
which is a lower bound to $F$~\cite{Miszczak2008}. Both enhanced bounds $T^{\;\star}_{\textrm{Sun}}$ and $T^{\;\star}_{\textrm{Deffner}}$ are larger than the respective bounds of Eqs.~\eqref{eq:sun}~and~\eqref{eq:deffner}. Therefore, whenever  $T_D$ is larger than the enhanced bounds it is also surely larger than the actual ones. Moreover, the enhanced bounds have orbit-independent terms that only depend on the following four parameters
\begin{align}
    \label{eq:x}
    &x: =\tr[\rho^2], \\
    \label{eq:y}
    &y:=\tr[\sigma^2], \\
    \label{eq:z}
    &z:=\tr[\rho\sigma], \\
    \label{eq:beta}
    &\beta:=\tr[\rho\sigma\rho\sigma],
\end{align}
where $x,y$ are bounded by $1/d$ from below and by $1$ from above, $z$ is bounded by $\sqrt{x y}$ from above, and $\beta$ is bounded by $z^2$ from above.
We proceed with the evaluation of the relative tightness of these bounds and of $T_D$ just by comparing their orbit-independent terms, obtaining
\begin{align}
    \label{eq:analytical_sun}
    &\bigg | 1 - \frac{z}{\sqrt{xy}} \bigg|\frac{\sqrt{x}}{2} \leq \sqrt{x+y-2z} \Rightarrow T^{\;\star}_{\textrm{Sun}} \leq T_D,
 \end{align}
and
\begin{gather}
    \bigg | 1 - \frac{z}{x} \bigg|x \leq \sqrt{x+y-2z}\Rightarrow T_{\textrm{Del Campo}} \leq T_D, 
 \end{gather}
for all $\rho,\sigma \in \mathcal{S}(\mathcal{H}_S)$ and all processes.

As mentioned in the main text, Deffner's bound is proven to be valid only when one of the two states is pure, i.e., for $\rho=\rho^2$ (or $\sigma=\sigma^2$)~\cite{Sun2015}, i.e., when $x=1$. Under this condition sub-fidelity, fidelity and super-fidelity all coincide~\cite{Miszczak2008} to be equal to $\sqrt{\tr[\rho\sigma]}$, and we obtain
\begin{gather}
\begin{split}
   &\sin^2\big[\arccos\big(\sqrt{z}\big)\big]\leq\sqrt{1+y-2z} \\
   &\Rightarrow T^\star_{\textrm{Deffner}}\leq T_D,
\end{split}
 \end{gather}
which proves our statement.
 
\section{Extending the validity of bound by Deffner et al.}
\label{a:deffner}
We will now show that our bound can be used to extend the validity of Deffner's bound~\cite{Deffner2013b} to the case of mixed initial states $\rho$, with $\tr[\rho^2]<1$.
As mentioned earlier, we can directly compare our bound $T_D$ to the enhanced bound $T^\star_{\textrm{Deffner}}$, given in Eq.~\eqref{eq:over_deffner}, which is always larger then the actual bound $T_{\textrm{Deffner}}$. Since our bound is valid for any initial state $\rho$, anytime $T_D$ is larger then $T^\star_{\textrm{Deffner}}$, then $T_{\textrm{Deffner}}$ is guaranteed to be valid for such values of $x$, $y$, $z$, and $\beta$.
Even though there is not a universal hierarchy between these two bounds, we can express a ranking between $T_D$ and $T^{\;\star}_{\textrm{Deffner}}$ using the following strategy:
We calculate the probability $p(T_D \geq T^*_{\textrm{Deffner}})$ of $T_D$ being larger than the upper bound on $T_{\textrm{Deffner}}$ in the space spanned by $z\in[0,\sqrt{x y}]$ and $\beta \in [0,z^2]$, as the ratio between the area where $T_D\geq T^*_{\textrm{Deffner}}$ and the area of the full space spanned by $z$ and $\beta$,
\begin{gather}
    \label{eq:probability_test}
    p(T_D\geq T^\star_{\textrm{Deffner}}) = \frac{\int_0^{\sqrt{xy}} \int_0^{z^2} \frac{\sgn(\Gamma (x,y,z,\theta))+1}{2} dz \;d\beta}{\int_0^{\sqrt{xy}} \int_0^{z^2} dz\; d\beta},
\end{gather}where $\sgn$ is the sign function, and
\begin{gather}
    \label{eq:performance_function}
    \begin{split}
    &\Gamma (x,y,z,\theta) = \sqrt{x+y-2z}\; + \\
    & \;\;- \sin^2\bigg[\arccos\bigg(\sqrt{z+\sqrt{2(z^2-\beta)}}\bigg)\bigg].
    \end{split}
\end{gather}

\begin{figure}
    \centering
    \includegraphics[width=0.49\textwidth]{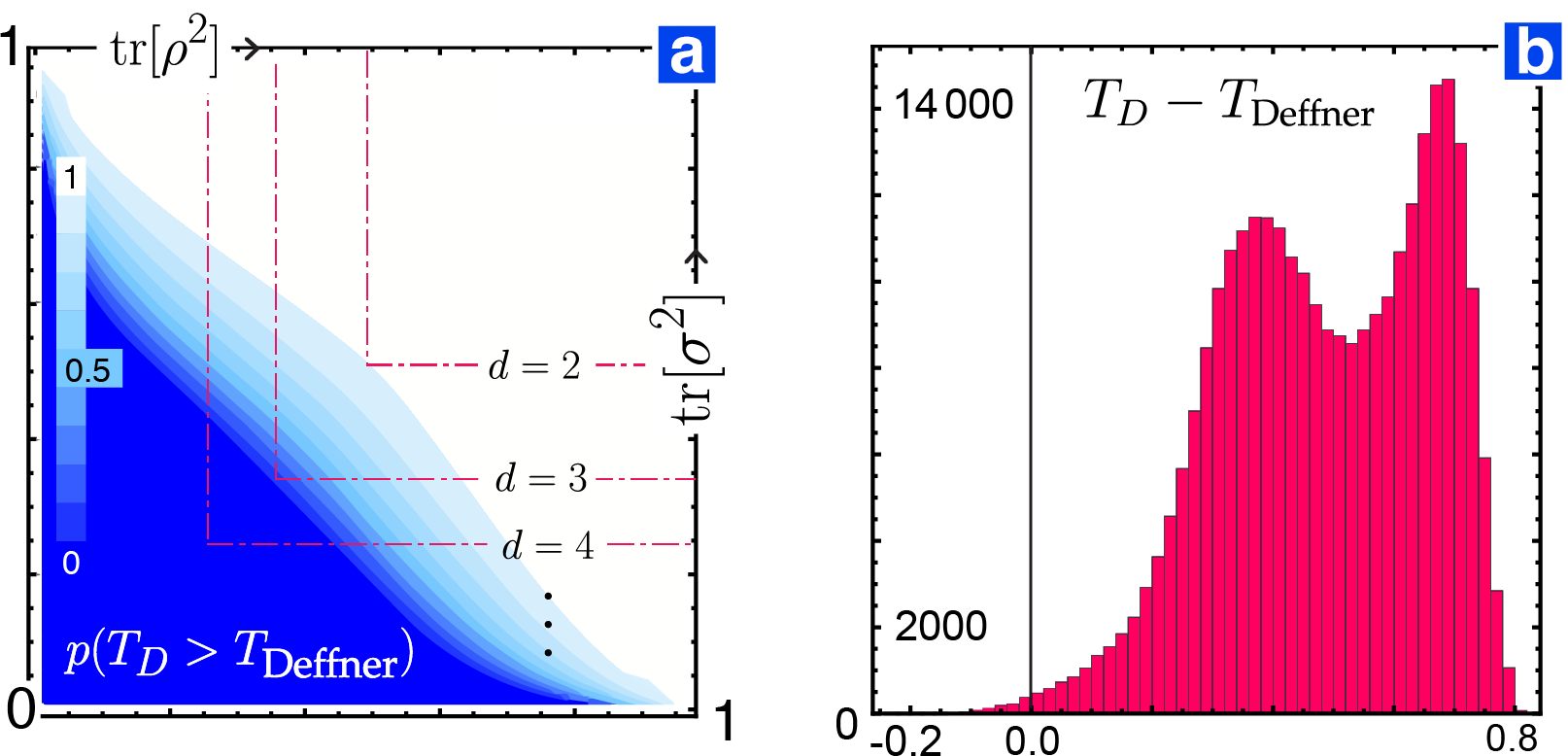}
    \caption{
    ({\fontfamily{phv}\selectfont\textbf{a}}) 
    The fraction of processes (quantified by their end points) for which $T_D$ outperforms an upper bound on $ T^\star_{\textrm{Deffner}}$, calculated as the ratio of the parameter space spanned by $\tr[\rho\sigma]$ and $\tr[\rho\sigma\rho\sigma]$ for which $T_D\geq T_{\textrm{Deffner}}$. A general \emph{rule of thumb} to evaluate the relative tightness between bound $T_D$ in Eq.~\eqref{eq:speed_limit_arbitrary} and Deffner's bound is given by $\tr[\sigma^2]\geq 1-\tr[\rho^2]\; \Rightarrow \;T_D\gtrsim T_{\textrm{Deffner}}$. Since $\tr[\rho^2],\tr[\sigma^2]\geq 1/d$, the region where $T_D$ outperforms Deffner's bound is always larger than that where the converse holds, as shown by the red dashed lines delimiting the physical region for $d=2,3,4$ (see Appending~\ref{a:deffner} for details).
    ({\fontfamily{phv}\selectfont \textbf{b}}) Numerical estimation of relative tightness between $T_D$ and $T_{\textrm{Deffner}}$, obtained sampling $3 \cdot 10^6$ initial and final states from the Bures and the Ginibre ensembles. Bound $T_D$ is almost always tighter than $T_{\textrm{Deffner}}$.}
    \label{fig:hierarchy_deffner}
\end{figure}

The probability $p(T_D\geq T^*_{\textrm{Deffner}})$ is a function of $x$ and $y$ measures how often $T_D$ is larger then Deffner in the space spanned by $z\in[0,\sqrt{xy}]$ and $\beta \in [0,z^2]$, given $x$ and $y$. As a result, we obtain a general \emph{rule of thumb} to decide which bound to use given the purity of initial and final states: For $y\geq 1- x$ bound $T_D$ is outperforms Deffner's (and vice versa for $y\leq 1-x$), as shown in Fig.~\ref{fig:hierarchy_deffner} {\fontfamily{phv}\selectfont \textbf{a}}.
Additionally, we have directly compared our bound $T_D$ to $T_{\textrm{Deffner}}$ numerically in Fig.~\ref{fig:hierarchy} { \fontfamily{phv}\selectfont \textbf{c}}, sampling $3\cdot10^6$ initial and final states from the Bures and the Ginebre ensembles. As can be seen from the figure, our bound outperforms Deffner's for the vast majority of the cases, as shown in Fig.~\ref{fig:hierarchy_deffner} {\fontfamily{phv}\selectfont \textbf{b}}.

\end{document}